\documentclass[runningheads]{llncs}

\usepackage{comment} 
\usepackage{float}
\usepackage{graphicx}
\usepackage[table]{xcolor}
\usepackage{multirow}
\usepackage{hyperref}
\usepackage{amsfonts}
\usepackage{ulem}
\usepackage{amsmath}

\providecommand{\ErdosRenyi}{Erd\H{o}s-R\'enyi }

\newcommand{\diam}{\text{diam}}
\newcommand{\den}{\text{den}}
\newcommand{\dist}{\text{dist}}
\usepackage[symbol]{footmisc}

\begin{document}
\title{
On Random Graph Properties}
%


\author{Hang Chen \and
Vahan Huroyan \and
Stephen Kobourov \and
Myroslav Kryven
}

\institute{
Department of Computer Science, University of Arizona}

\authorrunning{H. Chen, V. Huroyan, S. Kobourov, M. Kryven}

%
 
%
\maketitle              

\begin{abstract}

We consider 15 properties of labeled random graphs that are of interest in the graph-theoretical and the graph mining literature, such as clustering coefficients, centrality measures, spectral radius, degree assortativity, treedepth, treewidth, etc.  
We the analyze relationships and correlations between these properties. 
Whereas for graphs on small number of vertices we can exactly compute the average values and range for each property of interest, this becomes infeasible for larger graphs. 
We show that graphs generated by the \ErdosRenyi graph generator with $p = 1/2$ model well the underlying space of all labeled graphs with fixed number of vertices. The later observation allows us to analyze properties and correlations between these properties for larger graphs. We then use linear and non-linear models to predict a given property based on the others and for each property, we find the most predictive subset. 
We experimentally show that pairs and triples of properties have high predictive power, making it possible to estimate computationally expensive to compute properties with ones for which there are efficient algorithms.

\end{abstract}







\section{Introduction}
\label{sec:intro}

Understanding the most descriptive graph properties and the relations between them is important in theory as well as in applications such as graph mining. Recent developments such as Graph Neural Networks (GNN)~\cite{scarselli2008graph,wu2019comprehensive} and graph anonymization~\cite{aggarwal2011hardness} require careful analysis of various graph properties. GNNs are a natural extension of the deep learning algorithms to the graph domain where the input of the algorithm is a graph instead of a vector. In GNNs it is important to have descriptive and characteristic properties associated with the graph, its vertices, and its edges. 
Graph anonymization aims to provide privacy protection when working with graph data. Graphs that arise in social media contain sensitive information and publishing them and  conducting research on them might be problematic. Summarizing these graphs by their properties and creating graphs with similar properties reduces these concerns.

A graph is a pair $G = (V, E)$, where $V$ is a set of vertices, and $E$ is a set of edges that connect pairs of vertices. 
The space of labeled graphs  becomes very large very fast as already for $n=24$,  the number of labeled graphs with $24$ vertices exceeds the number of atoms in the universe ($10^{78}$).  To see how fast the set of labeled graphs grows, consider the number of labeled graphs for $n = 1, 2, \dots, 10$. Let $N_k$ be the number of labeled graphs with $k$ vertices, then $N_k = 2^{\binom{k}{2}}$: $N_1 = 1,$ $N_2 = 2,$ $N_3 = 8,$ $N_4 = 64,$ $N_5 = 1024,$ $N_6 = 32768,$ $N_7 = 2097152,$ $N_8 = 268435456,$ $N_9 = 6.8719e+10$, and $N_{10} = 3.5184e+13$. 
With this in mind, we study random graphs generated by the \ErdosRenyi model, which models well the underlying space of graphs. 
The \ErdosRenyi random graph generator takes two parameters: the number of vertices $|V|$ and the probability of selecting any edge $0 \le p \le 1$.

Some natural questions that arise include: What range of values do graph properties take, and are some of them correlated? Can we use a subset of properties to predict other properties? What are the most important properties for such predictions? 
However, summarizing a graph using a fixed number of properties can be misleading, as it is possible to have multiple different graphs with exactly the same values for each of the properties under consideration (even discounting isomorphic copies); see Chen et al.~\cite{chen2019same,chen2018same}. Thus, it is important to consider properties that are as descriptive as possible. With this in mind, we added more graph properties to original list of Chen et al.~and removed some that we have seen are highly correlated with the rest. 
Most importantly, in this work we consider labeled graphs while Chen et al.~\cite{chen2019same,chen2018same} considered the set of unlabeled (non-isomorphic) graphs.

The main motivation behind learning the relationships between graph properties is to find connections between them and be able to estimate some of the graph properties (especially the ones that are expensive to compute) based on the others. 
We  discuss two basic models for property  prediction 
which can be useful for larger graphs.
We also discuss an application, where we use the discussed properties to classify between different graph generators.

\subsection{Related Work}
There is a great deal of related work on graph mining, exploration of graph properties and graph generators with  
applications in bioinformatics, chemistry, software engineering, and social science. 
These properties range from basic, e.g., vertex count and edge count, to complex, e.g., clustering coefficients~\cite{kairam2012graphprism,li2011graph,mislove2007measurement} and average path length~\cite{chakrabarti2006graph,chakrabarti2007graph,mislove2007measurement}. They are widely used in graph mining applications and each captures and represents some important information about graphs. The node and edge connectivity may be used to describe the resilience of graphs~\cite{cartwright1956structural,loguinov2003graph}. Another commonly used graph property is the degree distribution. Many real-world graphs, including communication, citation, biological and social graphs follow a power-law shaped degree distribution
\cite{boccaletti2006complex,chakrabarti2006graph,newman2003structure}. Other real-world graphs have been found to follow an exponential degree distribution \cite{guimera2003self,sen2003small,wei2009worldwide}. Degree assortativity is of a particular interest in the study of social graphs and is calculated based on the Pearson correlation between the vertex degrees of connected pairs~\cite{newman2002assortative}.


Graph anonymization~\cite{aggarwal2011hardness,sun2013privacy} is another motivation for studying graph properties. Examples of such algorithms include $k$-neighborhood anonymity, edge randomization and cluster based generalization; see survey by Wu et al.~\cite{wu2010survey}. Another example includes Ying et al.~\cite{ying2008randomizing}, where the goal is to preserve the spectral information of the underlying graph.

Recently Chen et al.~\cite{chen2019same,chen2018same} consider different graph generators and the question of whether graph generators can represent and cover the space of \textit{non-isomorphic (unlabeled) graphs}. Experimental results show that no graph generator can model the underlying space of  non-isomorphic graphs well. However, as we show here, if isomorphism is allowed, then the \ErdosRenyi random graph generator does model the space of all labeled graphs well.

\textbf{Observation:}
To  argue the claim that the \ErdosRenyi random graph generator models the space of labeled graphs well, note that for a fixed $|V|=n$ and $p = 1/2$, the probability of the \ErdosRenyi graph generator selecting any labeled graph on $n$ vertices is equal to $1 / {\binom{|V|}{2}}.$ 
Therefore, graphs generated by the \ErdosRenyi model sample uniformly at random from the space of labeled graphs with fixed number of vertices.

\subsection{Our Contribution}
\label{sec:our_contribution}

First, we experimentally demonstrate that in terms of the graph properties considered in this work, the \ErdosRenyi graph generator with $p = 1/2$ models well the space of labeled graphs; see Figs.~\ref{fig:violin_conv_3} and \ref{fig:violin_conv_4} in the appendix. We observe that if one generates enough graphs for fixed number of vertices, there are many graphs that have the same exact set of properties but are different. In Fig.~\ref{fig:same_stat_pair_diff_v} in the appendix we report the number of times one needs to generate graphs to find a pair of different graphs with the same statistics, we report four such graphs in the appendix Section~\ref{sec:stability_tests}. 
We emphasise that in this work we focus on studying the relationships between graph properties for {\it labeled graphs} whereas Chen et al.~\cite{chen2019same} studied the properties of {\it unlabeled} graphs.
Second, we study the relationship between the  properties of interest and observe clusters of correlated properties that carry similar information. Different groups carry different information about the graphs.
Third, we consider the problem of predicting values of some graph properties based on the others and study the importance of each property on this prediction task. 
In particular, we study how well can we predict properties that are NP-hard to compute such as treewidth, treedepth, and vertex cover using properties that we can compute efficiently.
Fourth, in the appendix we discuss an  application to classify graphs that originate from different graph generators, based on properties of interest. We observe that a handful of properties from different groups suffice for accurate classification.


\section{Graph Properties}
\label{sec:def_properties}

In this section we consider different graph properties of interest in graph mining, bioinformatics, social science, and chemistry. The goal is to come up with a collection of descriptive graph properties, so that each graph can be uniquely (or almost uniquely) represented as a vector of its properties.
Since different fields use different graph properties, we include the ones that are frequently used in practice, 
varying from simple graph measures such as density, diameter and edge connectivity to more complex ones such as degree associativity and centrality measures; see Table.~\ref{table:properties}. In addition, we study several properties that are relevant for tackling intractable problems such as computing treewidth, treedepth, and vertex cover that we define below. These properties carry essential structural information about graphs, as many NP-hard problems become tractable on graphs where these properties are bounded~\cite{cygan2015parameterized}.

\begin{table}[t]

\caption{The set of graph properties considered in this paper. The first column includes the name of the property. The second column shows the formula (or the name of a solver) by which these properties can be computed for given graphs. The third column presents the time complexity of calculating the property and the last column includes relevant references.} 
\vspace{0.3cm}
\label{table:properties}
{
\renewcommand{\arraystretch}{1.5}%
\begin{tabular}{ m{3.2cm} m{6cm} m{1.4cm} m{1.2cm} } 

\hline\hline 
  Name & Formula & time & Reference\\ 
\hline 

\scalebox{0.75}[1]{Global Clustering Coeff.} 
& $GCC(G) =\frac{ 3 \times |\text{triangles}|}{|\text{connected triples in the graph}|}$& \scalebox{0.65}[1]{$O(|V|^3)$}&~\cite{chakrabarti2006graph,kairam2012graphprism}  \\

\scalebox{0.75}[1]{Average Sq. Clustering Coeff.}& \scalebox{0.85}[1]{$ASCC(G) =  \frac{1}{n} \sum_{i=1}^n c_4(u_i), u_i \in V,  n = |V| $} &\scalebox{0.65}[1]{$O(|V|^4)$}& \multirow{2}{*} ~\cite{lind2005cycles} \\  & $c_4(u_i) = \frac{\sum_{u=1}^{k_v}\sum_{w=u+1}^{k_v}q_v(u,w)}{\sum_{u=1}^{k_v}\sum_{w=u+1}^{k_v}[a_v(u,w)+q_v(u,w)] }$&\\

\scalebox{0.75}[1]{Average Path Length}& $APL(G) = \frac{1}{n(n-1)}\sum_{u,v \in V}d(u,v)$&\scalebox{0.65}[1]{$O(|V|^3)$}&~\cite{chakrabarti2007graph,li2011graph,chakrabarti2006graph,mislove2007measurement}\\

\scalebox{0.75}[1]{Degree Assortativity}& $r(G) = \frac{\sum_{xy}xy(e_{xy}-a_xb_y) }{\sigma_a \sigma_b}$&\scalebox{0.65}[1]{$O(|V|+|E|)$}&~\cite{newman2003mixing,mislove2007measurement}\\

\scalebox{0.75}[1]{Density} & $\den = \frac{2|E|}{|V|(|V| - 1)}$ & \scalebox{0.65}[1]{$O(1)$}&\\

\scalebox{0.75}[1]{Diameter}& $\diam(G) = \max\{\dist(v,w), v,w \in V \}$ &\scalebox{0.65}[1]{$O(|V||E|)$}& ~\cite{chakrabarti2007graph,mcglohon2011statistical,kairam2012graphprism,mislove2007measurement}\\

\scalebox{0.75}[1]{Edge Connectivity} & Ce: the minimum number of edges \shortstack{\\ to remove to disconnect the graph} &\scalebox{0.65}[1]{$O(|V||E|)$}& ~\cite{even1975network}\\


\scalebox{0.75}[1]{Closeness Centrality} & $C_C(G)$; see~\eqref{eq:freeman_cent}. $C_C(G, v) = \frac{n-1}{\sum_{u\neq v}d(u,v)}  $ &\scalebox{0.65}[1]{$O(|V||E|)$}& ~\cite{wasserman1994social,newman2018networks}\\

\scalebox{0.75}[1]{Betweenness Centrality}& $C_B(G)$; see~\eqref{eq:freeman_cent}. $C_B(G, v) =\sum_{s,t \in V} \frac{\sigma(s,t|v)}{\sigma(s,t)}$ &\scalebox{0.65}[1]{$O(|V|^2 \log(|V|))$}
& ~\cite{barthelemy2004betweenness,brandes2001faster,brandes2008variants,leydesdorff2007betweenness,newman2018networks}\\
\scalebox{0.75}[1]{Eigenvector Centrality }&$C_{Eig}(G)$; see~\eqref{eq:freeman_cent}. $A$ is the adjacency matrix and $C_{Eig}(G, v) = \sum_{u\in V} A_{v,u} C_{Eig}(G, u)$  &\scalebox{0.65}[1]{$O(|V|^3)$}& ~\cite{bonacich1987power,costenbader2003stability,borgatti2005centrality,lohmann2010eigenvector,newman2018networks}\\


\scalebox{0.75}[1]{Effective Graph Resistance}&$R_G = |V|\sum_{k=1}^{|V|-1}\frac{1}{\mu_k}$, 
&\scalebox{0.65}[1]{$O(|V|^3)$}
&~\cite{li2011graph,ellens2011effective}\\

\scalebox{0.75}[1]{Spectral Radius}&$\rho(G) = |\lambda_1|$
&\scalebox{0.65}[1]{$O(|V|^3)$}
&~\cite{li2011graph,berman2001spectral,friedland1988bounds}\\

\scalebox{0.75}[1]{Treewidth}
& \textsc{tidlib}; see Sec.~\ref{sec:def_properties}
&NP-Hard
&~\cite{bodlaender2005discovering,bodlaender2010treewidth,pace-tw}\\

\scalebox{0.75}[1]{Treedepth}
& \textsc{Bute-plus}; see Sec.~\ref{sec:def_properties}
&NP-Hard
&~\cite{pace-td}\\

\scalebox{0.75}[1]{Vertex cover}
& \textsc{WeGotYouCovered}; see Sec.~\ref{sec:def_properties}
&NP-Hard
&~\cite{pace-vc}\\

\hline 
\end{tabular}}
\end{table}

Note that some of the graph properties discussed below are only defined for connected graphs. With this in mind, we disregard the disconnected case in our analysis. It is known that for fixed $p$ and increasing values of $|V|$, the \ErdosRenyi model almost surely produces connected graphs~\cite{newman2018networks}. 
We experimentally confirm this by generating $10,000$ graphs for $|V| = 5, 6, \dots 15$ and examining the percentage of connected graphs; see Table.~\ref{table:connected_percentage}. 
For $p=1/2$ we get 99\% connected graphs for 13 or more vertices.

\begin{table}[t]
\caption{The  percentage of connected graphs from a set of $10,000$ generated by the ER model with $p=1/2$ and $p=\log(|V|)/|V|$ for increasing values of $|V|$.}
\vspace{0.3cm}
\begin{tabular}{|l|l|l|l|l|l|l|l|l|l|l|l|l|l|l|l|l|}
\hline
 & 5      & 6      & 7      & 8      & 9      & 10     & 11     & 12     & 13     & 14     & 15     
\\ \hline
 \scalebox{0.75}[1]{$p=1/2$}                                   & 59.8\% & 71.3\% & 81.6\% & 89.1\% & 93.7\% & 96.4\% & 98.1\% & 98.9\% & 99.4\% & 99.7\% & 99.8\% 
\\ \hline
 \scalebox{0.75}[1]{$p=\log(|V|)/|V|$}                              & 29.9\% & 30.6\% & 31.3\% & 32.0\% & 33.3\% & 33.8\% & 34.5\% & 35.4\% & 36.1\% & 36.4\% & 37.1\% 
\\ \hline
\end{tabular}

\label{table:connected_percentage}
\end{table}

The Global Clustering Coefficient (GCC)~\cite{chakrabarti2006graph,kairam2012graphprism} measures the tendency of a graph's vertices to cluster together. It computes the ratio of closed vertex triplets over all possible triplets.
The Average Square Clustering Coefficient (ASCC)~\cite{lind2005cycles} computes the ratio of closed vertex quadruples over all possible quadruples for each vertex. 
The Average Path Length (APL)~\cite{chakrabarti2007graph,li2011graph,chakrabarti2006graph,mislove2007measurement} measures the average of all shortest paths in the graph. The Degree Assortativity ($r$) measures whether vertices with high degrees are connected to other vertices with high degrees or not \cite{newman2003mixing,mislove2007measurement}, taking values between $-1$ and $1$. 
The graph Density (den)~\cite{even1975network} is the ratio between the number of edges of the graph and the maximum number of possible edges. The Diameter (diam) of a graph~\cite{chakrabarti2007graph,mcglohon2011statistical,kairam2012graphprism,mislove2007measurement} measures the greatest distance between any pair of vertices. Edge Connectivity ($C_e$)~\cite{even1975network} measures the minimum number of edges that need to be removed to disconnect the graph. Edge connectivity captures the robustness of the graph: in sparse graphs edge connectivity can vary whereas in dense graphs the variation decreases. 

Next we consider three centrality measues: closeness centrality, betweenness centrality and eigenvector centrality~\cite{newman2018networks}. All these are vertex-based measures but they can be interpreted as graph-based measures via centralization. We use Freeman centralization which measures unevenness~\cite{freeman1978centrality,ruhnau2000eigenvector}, where high/low centralization values represent high/low unevenness. 
Freeman's centrality is computed according to the following formula:
\begin{equation}
\label{eq:freeman_cent}
C_x(G) = \frac{\sum_{v\in V}(C_x(G, n^*)-C_x(G, v))}{\max_{G'}\sum_{v\in V'}(C_x(G', n^*)-C_x(G', v))},
\end{equation}
where $C_x(G, n^*)$ corresponds to the centralization of a node $n^*$ in graph $G$ and $x$ is the type of centralization ($C$ for closeness, $B$ for betweenness and $Eig$ for eigenvector). 
We remark that the denominator of \eqref{eq:freeman_cent} corresponds to the maximum possible value of the sum for all possible graphs $G$ with fixed number of vertices, effectively normalizing the values to the range $[0,1]$. 

The closeness centrality~\cite{wasserman1994social,newman2018networks} for a vertex measures the inverse of the average distance from the vertex to all others, describing how close vertices are to the center of the graph. 
Betweenness centrality~\cite{barthelemy2004betweenness,brandes2001faster,brandes2008variants,leydesdorff2007betweenness,newman2018networks} measures the influence of a vertex over the flow of information between every pair of vertices, assuming information flows over shortest paths between vertices. Eigenvector centrality~\cite{bonacich1987power,costenbader2003stability,borgatti2005centrality,lohmann2010eigenvector,newman2018networks} 
is a natural extension of  degree centrality and high eigenvector centrality means that a vertex is connected to many vertices with high eigenvector centrality values. 

The next two properties of interest are spectral radius~\cite{meghanathan2014spectral} and effective resistance~\cite{ellens2011effective}. The Spectral Radius ($R_G$) of a graph  is defined as the spectral radius of the corresponding adjacency matrix, given by the largest absolute value of its eigenvalues. The Effective Resistance ($R_G$) of a graph is defined as the sum of the effective resistances over all pairs of vertices, where the notion of resistance can be calculated by Ohm's law, when treating the graph as an electrical circuit.

We next consider treewidth, treedepth, and vertex cover. 
Even though computing these properties exactly is NP-hard,  they are important from an algorithmic point of view because the structure of graphs
where these properties are bounded is well understood and many NP-hard problems become tractable on such graphs~\cite{cygan2015parameterized}. For these properties we only provide intuitive explanations, as some of them are quite complicated and precise definitions are not  needed in this paper (but are provided in the papers cited when we introduce them below).


The first NP-hard to compute property is the
{\it treewidth (TW)}~\cite{bodlaender2005discovering,bodlaender2010treewidth} of an undirected graph, which measures how close the graph is to a tree: the smaller the treewidth, the closer the graph is to a tree. 
Graphs with bounded treewidth have useful structural properties, such as the existence of small graph separators. 
Dynamic programming can be applied on graphs with bounded treewidth to efficiently solve many NP-hard problems. In addition, Courcelle's theorem~\cite{courcelle2012book} can yield efficient algorithms for testing graph properties expressible in Monadic Second Order Logic, when a graph has bounded treewidth.

The second NP-hard to compute property is  {\it treedepth (TD)}~\cite{no-sgsa-2012}: 
the smallest possible height of a tree on the vertices of $G$ among all trees so that each edge of $G$ has an ancestor-descendant relationship in the tree. Graphs with bounded treedepth have useful structural properties, such as bounded longest path. Iterative pruning techniques, where irrelevant parts of the input are identified and removed to reduce the instance, have been successfully applied to graphs
with bounded treedepth to efficiently solve NP-hard problems~\cite{bhore-2019,no-sgsa-2012}.

The third NP-hard to compute property is the {\it vertex cover number  (VC)} of a graph, which is the smallest number of vertices that are incident to (cover) all the edges of the graph. Vertex cover is one of the original 21 NP-hard problems~\cite{karp}, and has been extensively in fixed parameter tractability~\cite{Fellows2018}.
In graphs with bounded vertex cover we can efficiently solve hard problems such as computing the exact crossing number of a graph, which is not known to be true for bounded treewidth or treedepth~\cite{hlinveny-2019}. Other NP-hard problems can also bee efficiently solvable for graphs with bounded vertex cover~\cite{bhore-2019,Bannister_2014}.

Given the utility of these three graph properties, and the fact that they are NP-hard to compute, it would be useful to predict them using other properties that can be efficiently computed. 
To train our prediction models we compute these properties exactly for the training set, using
state of the art implementations from the Parameterized Algorithms and Computational Experiments (PACE) Challenge. For treewidth we use the solver \textsc{tidlib} of Lukas Larisch, the winner of  PACE~2017~\cite{pace-tw} and the algorithm of choice in the algorithm selection study by Slavchev et al.~\cite{ml-tw}; for treedepth we use the solver \textsc{Bute-plus} of James Trimble, the winner of PACE~2020~\cite{pace-td};  and for vertex cover we use the sovler \textsc{WeGotYouCovered} of Hespe et al., the winner of PACE~2019~\cite{pace-vc}.



We remark that some of the graph properties are naturally bounded between $0$ and $1$ (see the centrality measures defined by~\eqref{eq:freeman_cent}) or $-1$ and $1$ (see assortativity).
For the sake of somewhat more uniform analysis, we linearly normalize all 
other properties 
so that their values lie between $0$ and $1$.

\section{Behavior of Graph Properties}
\label{sec:property_behavior}

\begin{figure}[htp!]
	\includegraphics[width=.24\textwidth]{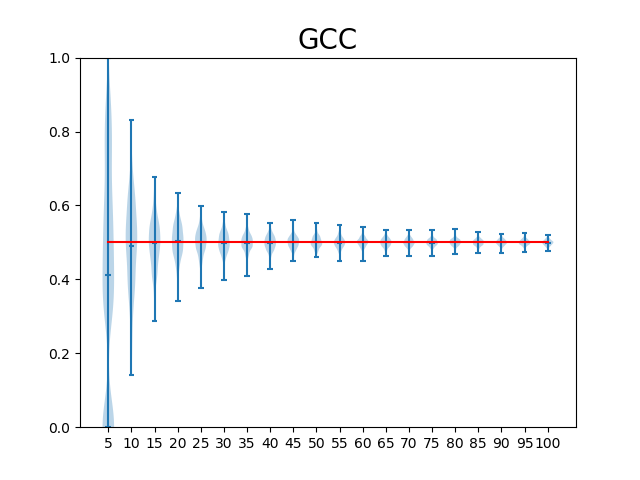}
	\includegraphics[width=.24\textwidth]{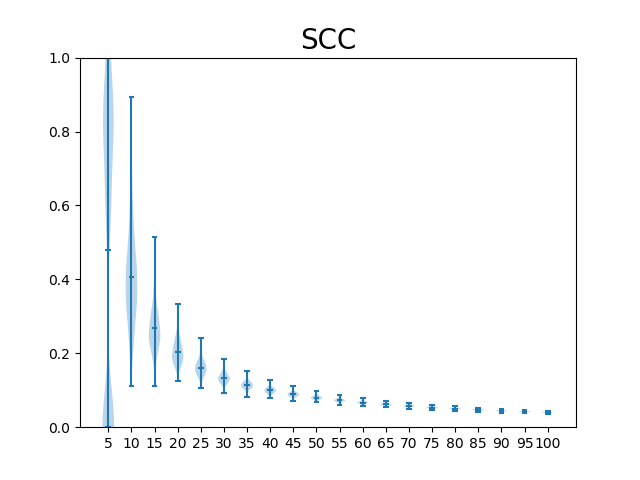}
    \includegraphics[width=.24\textwidth]{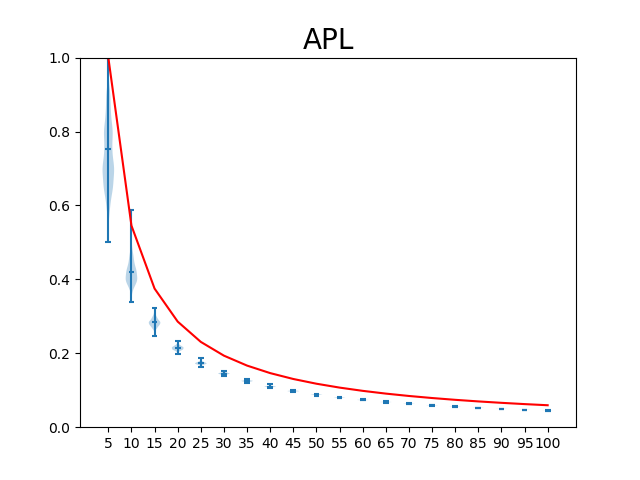}
	\includegraphics[width=.24\textwidth]{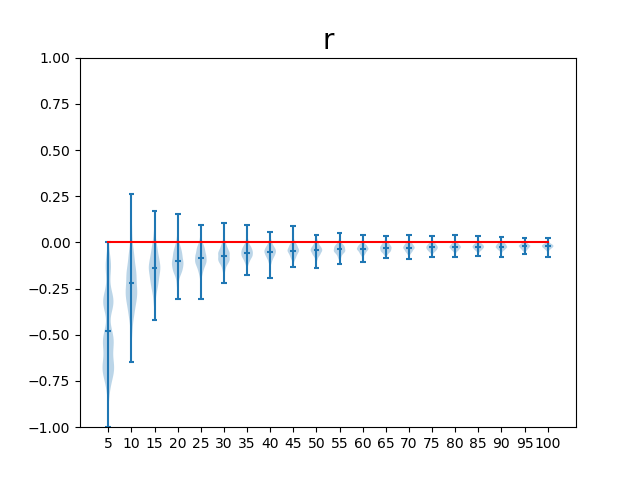}
	
	\includegraphics[width=.24\textwidth]{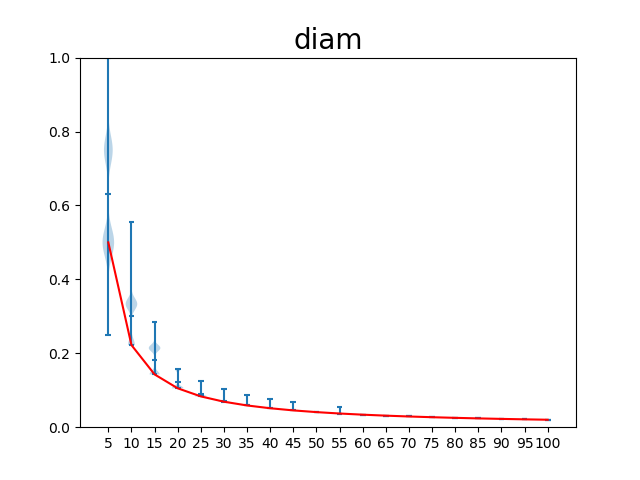}
	\includegraphics[width=.24\textwidth]{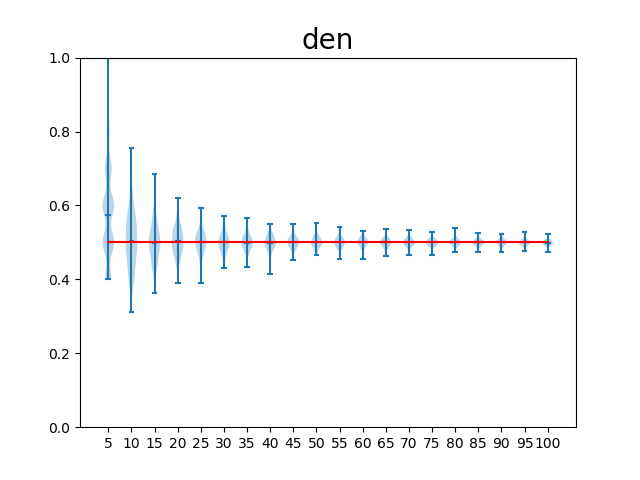}
	\includegraphics[width=.24\textwidth]{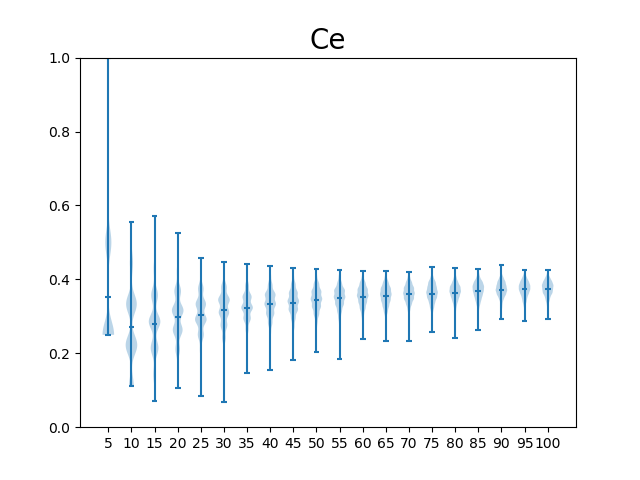}
	\includegraphics[width=.24\textwidth]{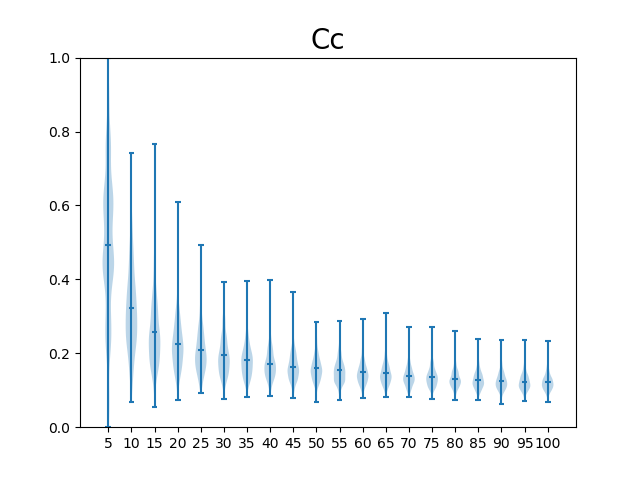}
	
	\includegraphics[width=.24\textwidth]{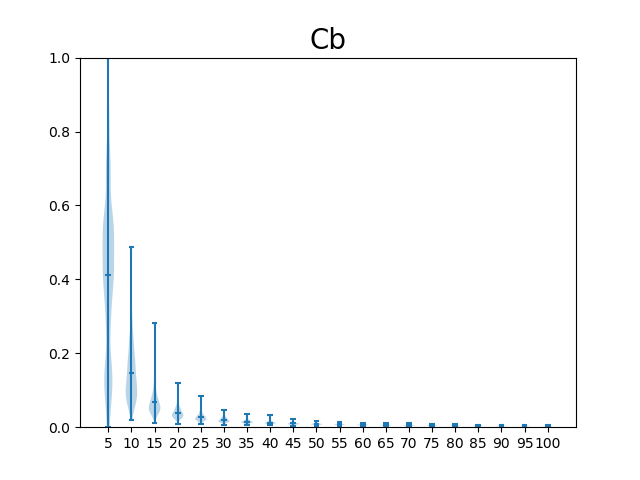}
    \includegraphics[width=.24\textwidth]{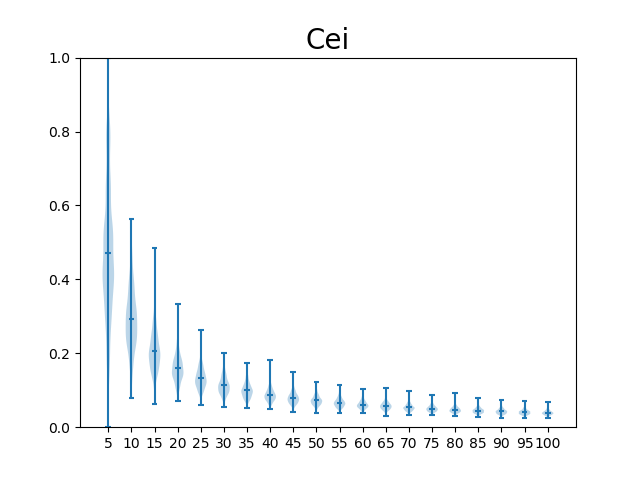}
    \includegraphics[width=.24\textwidth]{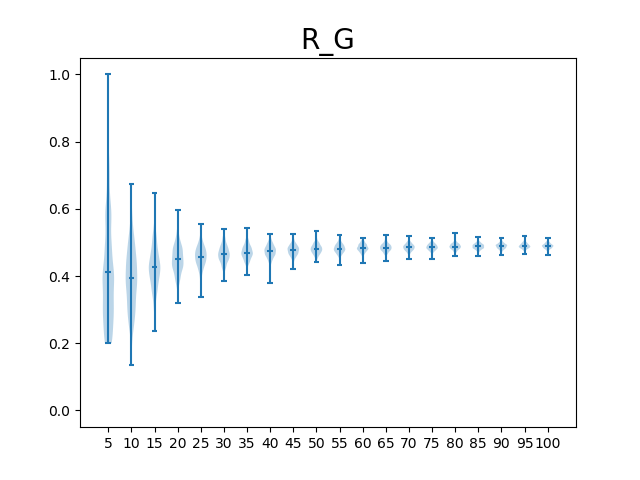}
    \includegraphics[width=.24\textwidth]{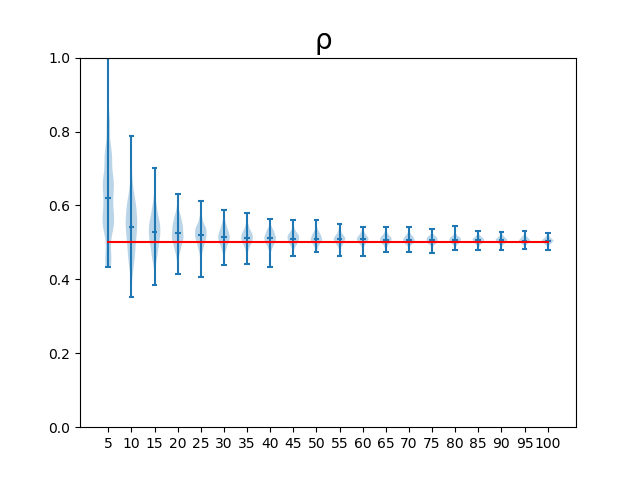}

    \includegraphics[width=.24\textwidth]{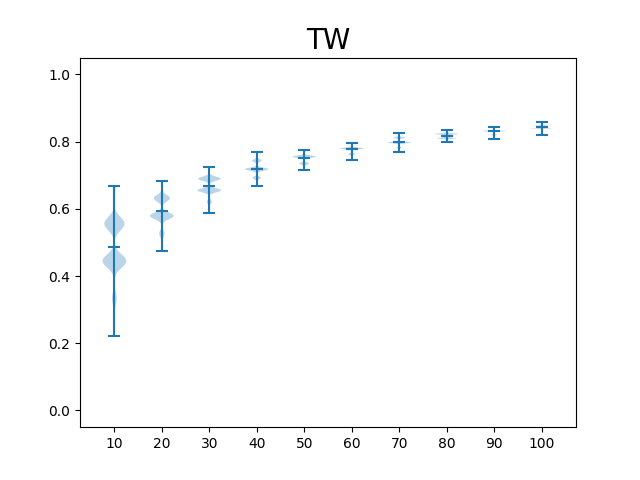}
    \includegraphics[width=.24\textwidth]{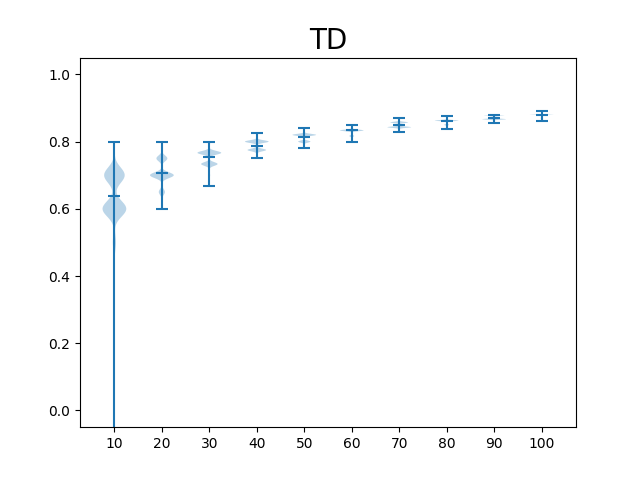}
    \includegraphics[width=.24\textwidth]{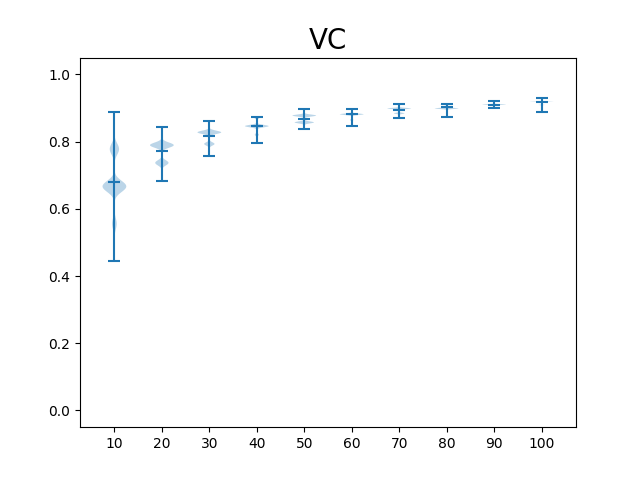}

\caption{Behavior of graph properties for $1000$ graphs generated according to \ErdosRenyi model with $p = 1/2$. The number of vertices is in the range $[5,10,..,100]$ and the results are illustrated with violin plots. The first row shows the results for GCC, ASCC, APL and $r$.
The second row includes diameter, density, $C_e$ and $C_C$. The third row includes $C_V$, $C_{Eig}$, $R_G$ and $\rho$. The fourth row includes TW, TD, and VC.
}
	\label{fig:violin_conv_05}
	\centering
\end{figure}

Here we consider the behavior of the graph properties of interest for graphs generated with the \ErdosRenyi model. The goal is to see how these properties change, with respect to the number of vertices in the graph. For some of the properties asymptotic results are known and proved for various values of $p$; see Table~\ref{Table:converage_props}. We study these results numerically,  and verify them against known asymptotic bounds. Note that for properties with unknown asymptotic bounds we also see a trend of convergence; see Fig~\ref{fig:violin_conv_05}. As the number of possible labeled graphs grows exponentially in the number of vertices, we generate graphs based on \ErdosRenyi model with $|V| = 5, 10, \dots, 100$ using the following values of $p$: $p = 1/2$ (Fig.~\ref{fig:violin_conv_05}) and $p = \log(|V|)/|V|$ (Fig.~\ref{fig:violin_conv_log}). 
For each experiment we generate $1000$ graphs and compute the $15$ properties. 
For the properties where asymptotic bounds are known we show these bounds by a red curve and we remark that our experimental results match  the theoretical bounds, which is a good indication that the experimental results for properties without theoretical bounds are plausible. We next discuss some of our findings.
\begin{table}[t]
\caption{Known bounds for some of the properties  \ErdosRenyi graphs. 
The first row shows the bounds for \ErdosRenyi graphs with $p = 1/2$ and the second row shows the bounds for \ErdosRenyi graphs with $p = \log (|V|) / |V|$. Note that the bounds of this table are for the non-normalized properties. Later, in Figure~\ref{fig:violin_conv_05} we show the normalized versions of these bounds.
}
\vspace{0.3cm}
\begin{tabular}{|l|l|l|l|l|l|l|l|}
\hline
& GCC     & APL    & r     & den      & diam  & $R_G$    & $\rho$ \\ \hline
\scalebox{0.75}[1]{$p=1/2$}  & $p$~\cite{newman2018networks} &  \scalebox{0.75}[1]{$\frac{\ln(|V|)}{\ln(|V|-1)p}$}~\cite{newman2018networks} & $0$~\cite{newman2002assortative} & $p$ & 2~\cite{blass1979properties} &  \scalebox{0.5}[1]{$\leq|V|-1+\frac{2}{|V|-2}$}~\cite{ellens2011effective} & \scalebox{0.65}[1]{$(1+o(1)max(\sqrt{\Delta}, np)$}~\cite{krivelevich2003largest}\\ \hline


\scalebox{0.65}[1]{$p=\log(|V|)/|V|$} & $p$~\cite{newman2018networks}  & 
\scalebox{0.65}[1]{$\approx\frac{\ln(|V|)}{\ln(|V|-1)p}$}~\cite{newman2018networks} 
& 0~\cite{newman2002assortative} & $p$  &
\scalebox{0.65}[1]{$\approx \frac{\log(|V|)}{\log(|V|p)}$}~\cite{chung2001diameter}   
&{$\approx \frac{|V|}{p}$}~\cite{sylvester2016random} & \scalebox{0.65}[1]{$(1+o(1)max(\sqrt{\Delta}, np)$}~\cite{krivelevich2003largest} \\ \hline
\end{tabular}

\label{Table:converage_props}
\end{table}

According to Fig.~\ref{fig:violin_conv_05} some of the properties quickly converge to certain values for both \ErdosRenyi with $p = 1/2$ and $p = \log (|V|) / |V|$. For example see the results for GCC, ASCC, APL, diameter, density, effective graph resistance. Others converge faster for one of the two models and slower for the other one. For example for \ErdosRenyi with $p = 1/2$ the diameter is almost always 2, thus the normalized version of the diameter converges to $0$ very quickly ($2/|V|$), while for \ErdosRenyi with $\log (|V|) / |V|$ it is not clear whether diameter converges to $0$ or not. Another experimental observation is that edge connectivity for \ErdosRenyi with $p=1/2$ converges slower than for \ErdosRenyi with $p=\log(|V|)/|V|$.
This might be due to the fact that graphs generated by \ErdosRenyi with $p=1/2$ are  denser  and density is strongly correlated with edge connectivity. 

We remark that the trends for TW, TD and VC are similar and although we are not aware of known convergence limits, experimentally they seem to converge.
Gao~\cite{gao-2009} showed that TW of \ErdosRenyi graphs for $p=1/2$ is, with
high probability, greater than $\beta |V|$ for some constant $\beta > 0$ if 
the edge/vertex ratio $\frac{|E|}{|V|}$ is greater than 1.073. Our experiments indicate that $\beta \approx 0.83$.

Finally, note that there are several graph properties (e.g., the  centrality measures) that do not have known bounds and it is not clear whether they converge.

\section{Property Correlation and Prediction}
\label{sec:property_relationships}

In this section we explore the correlations between graph properties for the set of all graphs ($|V| \le 7$) and then for graphs generated by \ErdosRenyi model. Next, we use linear and non-linear models to predict some graph properties based on the others. At the end we apply some feature selection techniques to understand which features are most important for predicting other properties.

\subsection{Exploring  correlations between  properties}
\begin{figure}
\centering
	\includegraphics[width=0.95\linewidth]{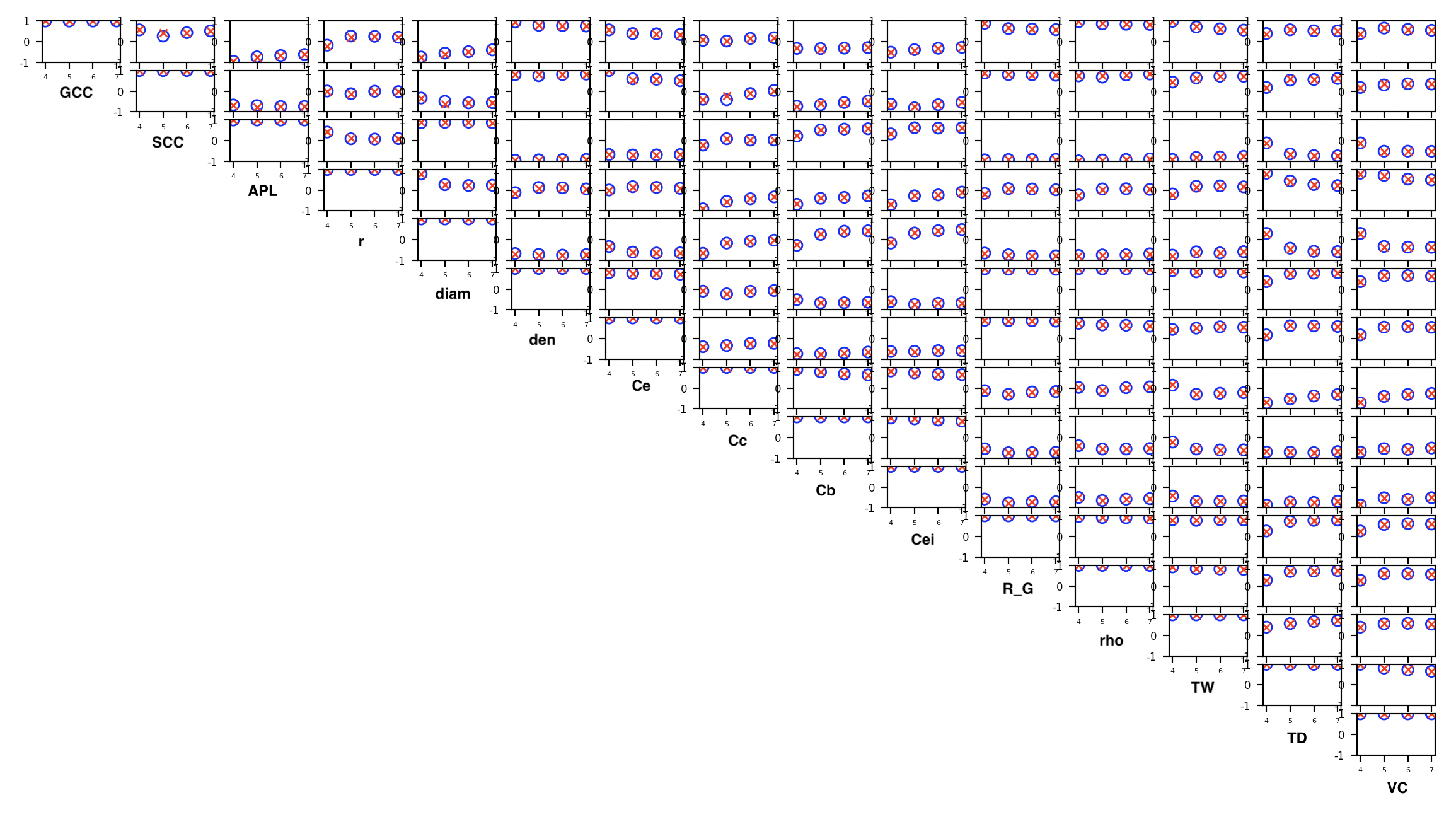}
	\caption{Comparison between correlations of graph properties for the set of all graphs (see the blue circles) and a sample of $1000$ graphs generated by the \ErdosRenyi model with $p = 1/2$ (see the red crosses) for $|V|=4,5,6,7$. For each scatter plot, the $x$-axis shows the value of $|V|$ and the $y$-axis is the correlation, which ranges from $-1$ to $1$.}
	\label{fig:sample_correlation_trend}
\end{figure}

We aim to study the correlations between the $15$ graph properties discussed in Section~\ref{sec:def_properties}. First, we compute the correlations between the properties for all graphs with $|V| = 4,5,6,7$. This analysis is special since we actually have the set of all labeled graphs for $|V| \le 7$. Thus, we can see whether the computed correlations for a sample of graphs generated by the \ErdosRenyi model match with the correlations of the set of all graphs. For each $|V| = 4, 5, 6, 7$ we generate $1000$ graphs according to \ErdosRenyi model with $p = 1/2$ and for every two properties we compute the correlation between them. We also compute these correlations for the set of all graphs with $|V| = 4, 5, 6, 7$. In Fig.~\ref{fig:sample_correlation_trend} we report the results. The blue circles correspond to the values for the set of all graphs and red crosses correspond to the values for the sample generated by the \ErdosRenyi model $p = 1/2$. According to 
Fig.~\ref{fig:sample_correlation_trend} the results match, which supports our further analysis, where for larger graphs we only consider graphs generated by the \ErdosRenyi model with $p = 1/2.$

A natural question to ask is, how big of a sample one should take to obtain comparable correlation results with that of the set of all graphs for larger values of $|V|.$ It is impossible to find an exact answer to this question, since as $|V|$ grows the set of all graphs with $|V|$ vertices grows too fast, exceeding the number of atoms in the universe ($\approx 10^{78}$) already for $|V|=24$. We propose to answer this question with the following stability test: For $|V| = 100$ we sample $100, 200, 400, 800$ and $1600$ graphs according to \ErdosRenyi model and compute the corresponding correlations between the properties. We repeat this experiment 10 times and report the results in Fig.~\ref{fig:stable_test_v100}. The idea is that once we start getting consistent results, i.e.,  the variation between the correlations is small, we can assume that these are the correct correlations between the properties for the total dataset. From Fig.~\ref{fig:stable_test_v100}, we can see the violin plot range diminishes with larger sample size and the results for $|S| = 1600$ are consistent.

\begin{figure}[t]
\centering
\includegraphics[width=\columnwidth,height=0.8cm]{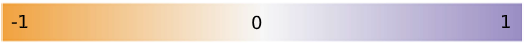}
\includegraphics[width = 0.45\textwidth]{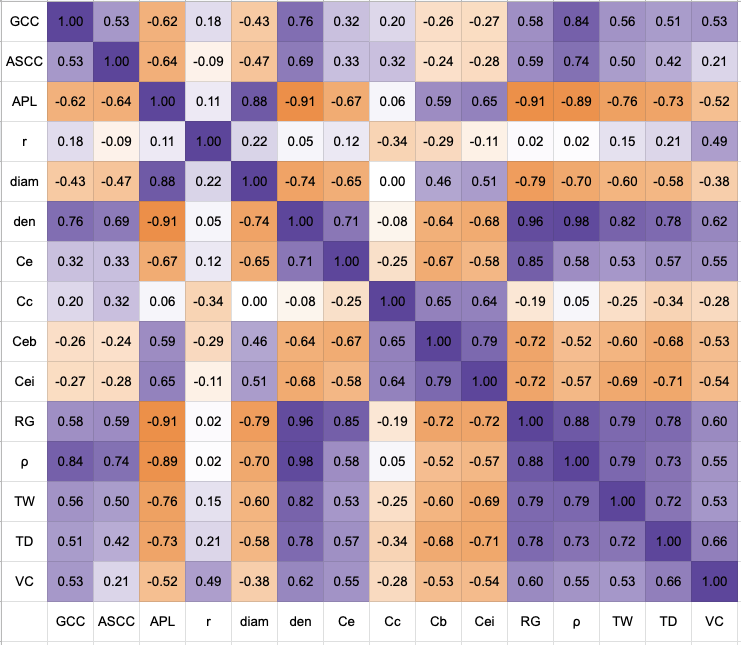}
\includegraphics[width = 0.45\textwidth]{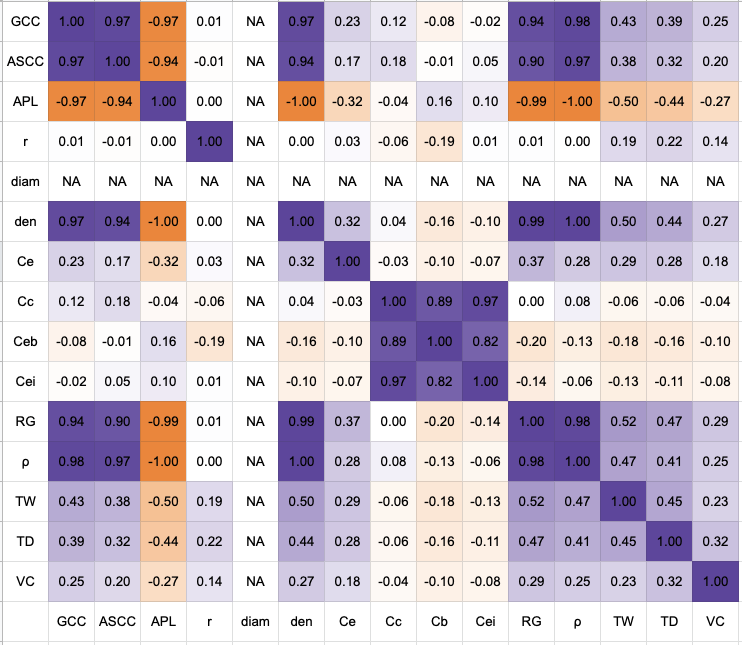}
\caption{Property correlations  for all isomorphic graphs with $|V| = 7$ (left), 
and 1000 graphs generated by the \ErdosRenyi model, $|V| = 100$, $p=1/2$ (right).
}
\label{fig:heatmap}
\end{figure}

We note that the correlation analysis varies with $|V|$. 
Thus, we compare the correlations between the properties for graphs with $|V| = 7$ and $|V| = 100$; see Fig.~\ref{fig:heatmap}\footnote{Here and in Fig.~\ref{fig:linear_model_corr} NA values are due to the fact that for large values of $n$ the diameter of graphs generated by \ErdosRenyi model, $p=1/2$ is $2$ with high probability. }
For $|V| = 100$, the weak correlations become weaker, while strong correlations become stronger.



\subsection{Linear and non-linear models for prediction}
\label{sec:linear_analysis}

Studying the correlations helps to understand how these $15$ properties are correlated for a given sample, although it does not fully answer the questions of how to use some of these properties to predict the rest. With this in mind, we propose a simple experiment: for a given property, we run a basic linear regression model to see whether we can predict it by the other $14$ properties. For this experiment we generate $32,000$ graphs by the \ErdosRenyi model with $p=1/2$ and $|V| = 100$. We compute the $15$ properties defined in Table~\ref{table:properties}. Next, we randomly separate this dataset into training, validation and test datasets with $80\%,10\%,10\%$ of the data, respectively. We use the basic linear regression model for each predictor. We learn the parameters of the linear regression model on the training dataset and compute the prediction accuracy error for the test dataset. As a baseline we take the mean predictor, that is the mean values of the training set. As a loss we use $L_1$ error, which is the $L_1$ distance between the predictor and the true value. 
The percentage error for the predicted values is shown in Table~\ref{table:prediction_feature}.


As we can see the linear model for all of the properties significantly improves compared to the baseline predictor. Comparing the linear predictor to the baseline mean predictor, $\rho$ 
(from $0.554\%$ to $0.004\%$) and $R_G$ (from $0.574\%$ to $0.006\%$) improve the most.

\begin{table}[t]
\centering
 \caption{Prediction error (in percentage) for the baseline mean estimator (the first row), the linear regression estimator (the second row) and the non-linear estimator, described in Section~\ref{sec:linear_analysis}. The columns correspond to the $15$ graph properties.}
\vspace{0.3cm}
\begin{tabular}{|l|l|l|l|l|l|l|l|l|l|l|l|l|l|l|l|}
\hline

 \scalebox{0.75}[1]{$L_1$ loss}       & \scalebox{0.75}[1]{GCC}     & \scalebox{0.75}[1]{ASCC}     & \scalebox{0.75}[1]{APL}     & \scalebox{0.75}[1]{r}       & \scalebox{0.75}[1]{diam}     & \scalebox{0.75}[1]{den}    & \scalebox{0.75}[1]{Ce}     & \scalebox{0.75}[1]{Cc}      & \scalebox{0.75}[1]{Cb}      & \scalebox{0.75}[1]{Cei}    & \scalebox{0.75}[1]{$R_G$}     & \scalebox{0.75}[1]{$\rho$} & \scalebox{0.75}[1]{TW}  & \scalebox{0.75}[1]{TD} & \scalebox{0.75}[1]{VC} \\ \hline
 
\scalebox{0.75}[1]{mean}     & \scalebox{0.75}[1]{0.575} & \scalebox{0.75}[1]{0.091} & \scalebox{0.75}[1]{0.017} & \scalebox{0.75}[1]{0.564} & \scalebox{0.75}[1]{0.000} & \scalebox{0.75}[1]{0.558}    & \scalebox{0.75}[1]{1.757} & \scalebox{0.75}[1]{1.745} & \scalebox{0.75}[1]{0.045} & \scalebox{0.75}[1]{0.493} & \scalebox{0.75}[1]{0.574} & \scalebox{0.75}[1]{0.554} & \scalebox{0.75}[1]{0.530} & \scalebox{0.75}[1]{0.206} & \scalebox{0.75}[1]{0.337} \\ \hline

\scalebox{0.75}[1]{linear}    & \scalebox{0.75}[1]{0.093} & \scalebox{0.75}[1]{0.003} & \scalebox{0.75}[1]{0.000*} & \scalebox{0.75}[1]{0.380} & \scalebox{0.75}[1]{0.000} & \scalebox{0.75}[1]{0.003}    & \scalebox{0.75}[1]{1.033} & \scalebox{0.75}[1]{0.198} & \scalebox{0.75}[1]{0.013} & \scalebox{0.75}[1]{0.076} & \scalebox{0.75}[1]{0.006} & \scalebox{0.75}[1]{0.004} & \scalebox{0.75}[1]{0.377} & \scalebox{0.75}[1]{0.283} & \scalebox{0.75}[1]{0.306} \\ \hline
 
\scalebox{0.75}[1]{non-linear} &\scalebox{0.75}[1]{0.092} & \scalebox{0.75}[1]{0.002} & \scalebox{0.75}[1]{0.000*} & \scalebox{0.75}[1]{0.334} & \scalebox{0.75}[1]{0.000} & \scalebox{0.75}[1]{0.002} & \scalebox{0.75}[1]{1.013} & \scalebox{0.75}[1]{0.190} & \scalebox{0.75}[1]{0.011} & \scalebox{0.75}[1]{0.070} & \scalebox{0.75}[1]{0.005} & \scalebox{0.75}[1]{0.003} & \scalebox{0.75}[1]{0.371} & 
\scalebox{0.75}[1]{0.266} & \scalebox{0.75}[1]{0.288} \\ \hline
\end{tabular}
\vspace{-.2cm}

\label{table:prediction_feature}
\end{table}

However, the linear model has its limitations as there are likely some non-linear connections between the properties. Thus, for the next experiment we add some non-linear combinations of the properties. We use all the second order combinations, the square roots and the logarithms of the properties. To avoid complications we use the absolute values for the square roots $\sqrt{|x|}$ and  we take $\log{(1+|x|)}$ for each of the property. After adding all these features we run another linear regression for this dataset; the results are shown in Table~\ref{table:prediction_feature}. We can see some significant improvements compared to the linear regression model. Comparing the non-linear to the linear predictor,
ASCC (from $0.03\%$ to $0.02\%$) $C_C$ (from $0.198\%$ to $0.190\%$) and $r$ (from $0.380\%$ to $0.334\%$) improve the most.

For the NP-hard properties, we can predict the values of TW with an error of $0.371\%$, TD with an error of $0.206\%$, and VC with an error of $0.288\%$, although these values are not consistently better than the baseline.

\subsection{Feature selection}
\label{sec:feature_selection}


\begin{figure}[hb!]
\centering
    \includegraphics[width=0.48\textwidth]{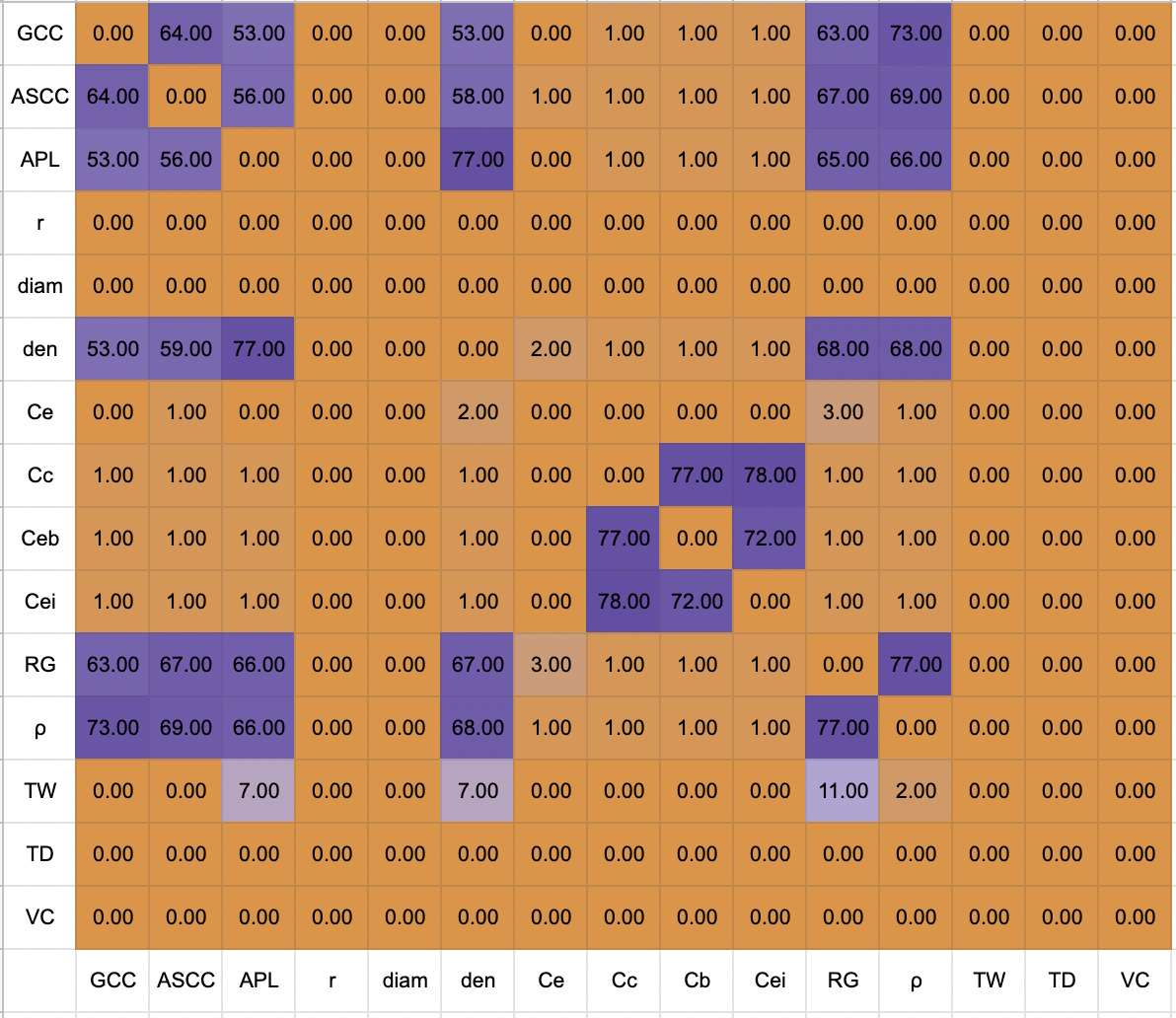}
    \includegraphics[width=0.48\textwidth]{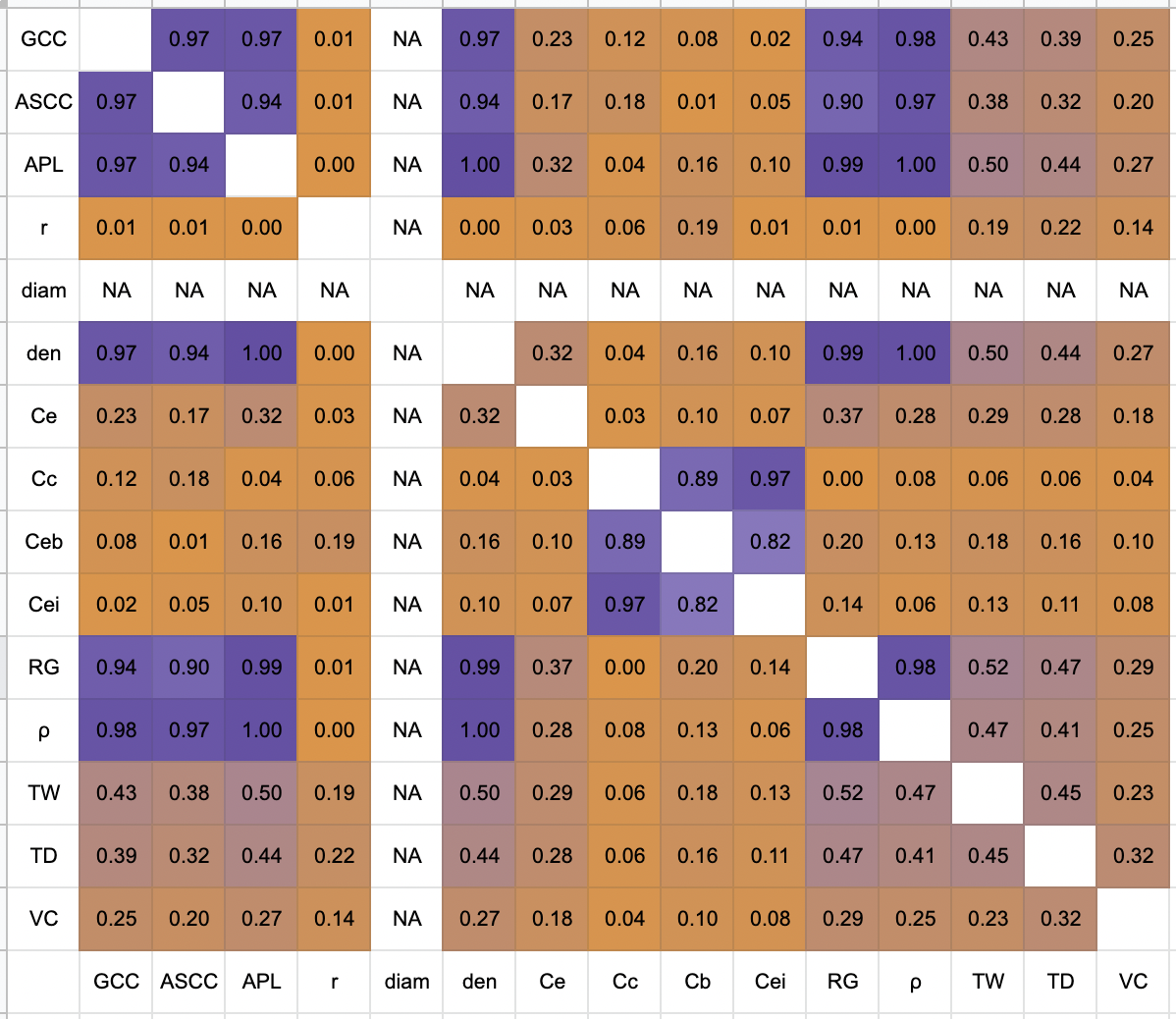}
    \caption{The property importance matrix (left);
    absolute values of the correlations for all pairs of properties for 1000 graphs generated by ER model, $p = 0.5$ (right).}
  \label{fig:linear_model_corr}
\end{figure}

The analysis in Section~\ref{sec:linear_analysis} leads to the question: which properties are important for the prediction of other properties? 
We propose the following feature selection technique: For each property, we fix $2$ other properties that we call \textit{predictors} and run a linear regression model on it. Next we pick another property and add it to the set of predictors. We run a linear regression with these 3 predictors and if the improvement on the loss is at least $20\%$ (we choose this threshold experimentally) we declare this property as an important predictor. We repeat this experiment for all possible properties and predictors and record the results in a matrix (the property importance matrix). Each cell of the property importance matrix records the number of times that the particular predictor has been important; see the left subfigure of Fig.~\ref{fig:linear_model_corr}. For comparison, the right subfigure of Fig.~\ref{fig:linear_model_corr} shows the matrix of absolute values of correlations between properties. We can see that the two are similar and thus our proposed technique seems to be a plausible alternative to the correlation computation.
Both figures show that density is a useful predictor for GCC, ASCC, effective graph resistance and spectral radius. This is useful as density is easy to compute compared to other graph properties.
We can also see that APL, density, and $R_G$ were helpful in predicting TW. 


The final experiment that we run aims to estimate TW, TD and VC based on the other 12 properties. We apply the forward subset selection for the $32000$ graphs generated by the \ErdosRenyi graph generator with $p = 1/2$ and best subset selection techniques to figure out the properties that are the most important for the estimation of TW, TD, and VC. We report the $R^2$ value for each model which tells us which percentage of the total variation the model was able to capture. 
Both methods have identical results, according to our experiments.

The most important properties to estimate TW are $R_G$, APL, $r$, and GCC in this order for models with one, two, three, and four estimator properties and the values of $R^2$ are $0.27467$, $0.32550$, $0.35155$, and $0.35277$ respectively.
For TD the important properties are $R_G$, density, $r$, and GCC in this order and the values of $R^2$ are $0.21772$, $0.27420$, $0.30997$, and $0.31469$ respectively.
Finally, for VC, the most important properties are $R_G$ density, $r$, and GCC and the values of $R^2$ are $0.08176$,
$0.10349$, $0.11873$, and $0.12912$  respectively.





\section{Conclusions and Future Work}

Our results indicate that \ErdosRenyi graphs can be used to study the properties of the space of all labeled graphs. 
Determining the natural dimension of the space of graphs (when treating each graph as a high dimensional vector based on its properties) seems a challenging but useful research direction. 
We also considered the problem of predicting expensive-to-compute graph properties. This is a challenging task as 
such properties describe a significantly richer structural information about graphs than other efficiently-computable properties. We see that linear and non-linear regression models can predict the values of treewidth, treedepth, and vertex cover with an error of at most $0.38\%$. This, however, is a slight improvement over the baseline, which might be due to the inherent computational complexity of these properties, or due to the subset of properties under consideration. 
%
%
Given the groups of correlated properties in our list of $15$, putting together a new list that captures more and more diverse information about graphs is a natural direction for future work.


\bibliographystyle{abbrv}
\bibliography{graph}

\newpage


\appendix

\noindent \Large{\textbf{Appendix}} \\

\noindent The appendix consists of two sections. Section~\ref{sec:stability_tests} provides details about the scalability of the correlation computations from Section 4. We numerically demonstrate that for graphs up to $100$ vertices \ErdosRenyi graph generator with $1000$ instances captures well the total set of graphs (in terms of correlations). Section~\ref{sec:graph_classications}  extends our analysis of graph properties to distinguish between different graph generators.  
The appendix concludes with additional figures, such as those for the \ErdosRenyi graph generator with $p = \log(|V|) / |V|$.

\section{Stability Tests}
\label{sec:stability_tests}
\begin{figure}[ht]
	\includegraphics[width=\textwidth]{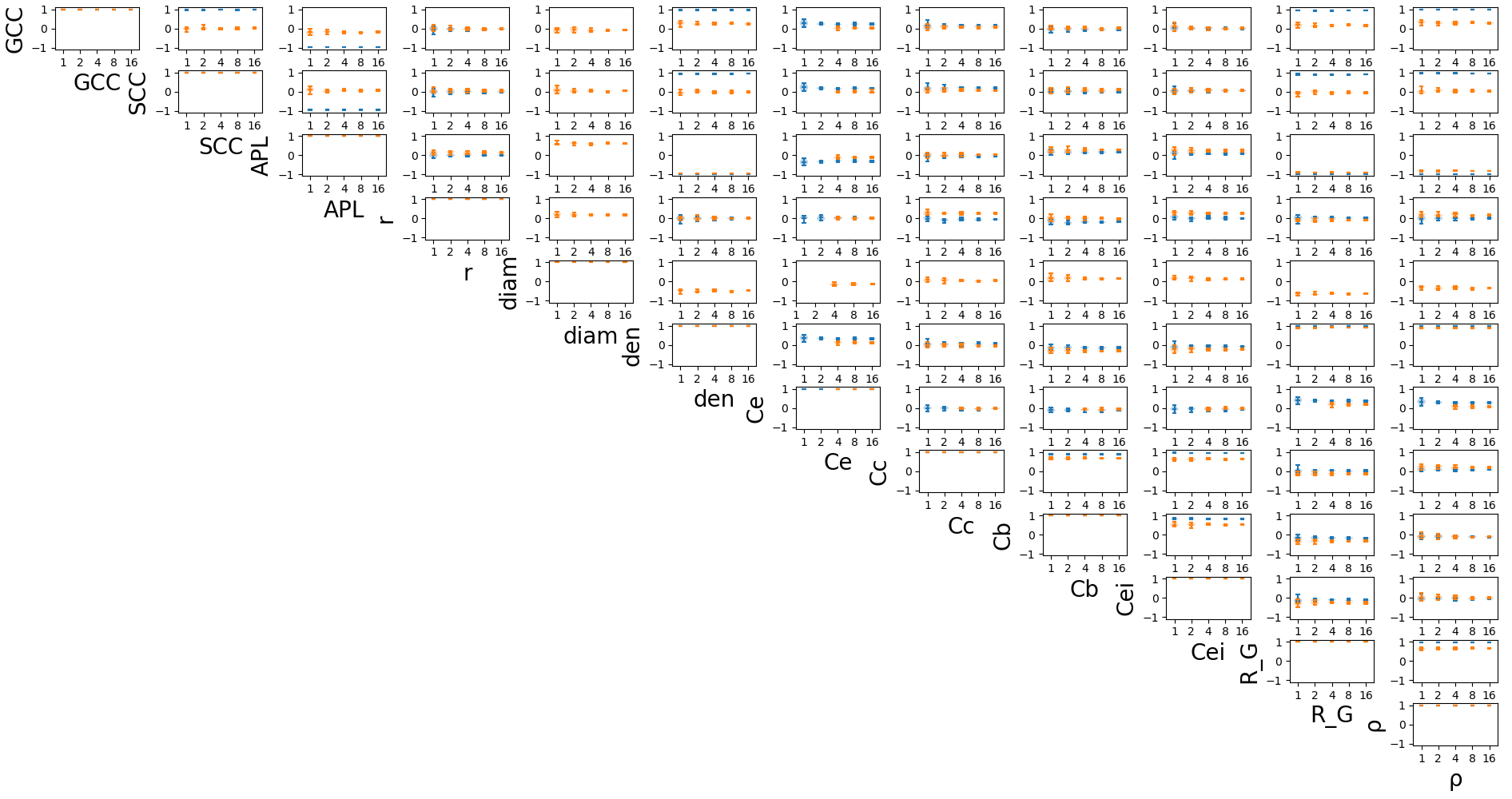}
	\caption{Stability of the correlations between the graph properties for \ErdosRenyi graphs with $p=1/2$ and $p=\log(|V|)/|V|$ for $|V| = 100$. For each model we generate $\{100, 200, 400, 800, 1600\}$ graphs and compute the correlations between the graph properties. We repeat this $10$ times and report the results as violin plots for $|S|=\{100,200,400,800,1600\}$ (where $|S|$ is the size of the sample).}
	\label{fig:stable_test_v100}
	\centering
\end{figure}

Fig.~\ref{fig:stable_test_v100}, demonstrates some stability tests for the \ErdosRenyi model with $p = 1/2$ and $p=\log(|V|)/|V|$. The idea is to observe how big of a sample one should take for various values of $V$ to obtain correlations. We create samples of different sizes and observe the variances between the correlation values. The smaller the variances are the more stable the samples are.

\begin{figure}[ht]
    \centering
	\includegraphics[width=.6\textwidth]{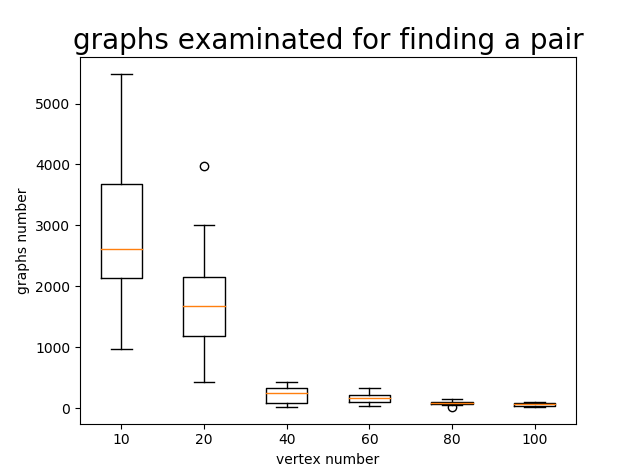}

	\caption{Number of graphs that need to be generated by the \ErdosRenyi model with $p = 1/2$ to find a pair of graphs which have the same exact set of $15$ graph properties up to 2 decimal places. We run this 10 times and report the results in boxplots.}
	\label{fig:same_stat_pair_diff_v}
	\centering
\end{figure}

In Fig.~\ref{fig:same_stat_pair_diff_v} we report the graphs that need to be generated to find 2 graphs with the same properties up to 2 decimal places. 
We remark that even though we are able to find such graphs, most of them have different degree distributions and thus are different graphs. We also remark that as the number of vertices increase one needs to generate less graphs to find different graphs with the same statistics.

\begin{figure}[ht!]
	\includegraphics[width=.24\textwidth]{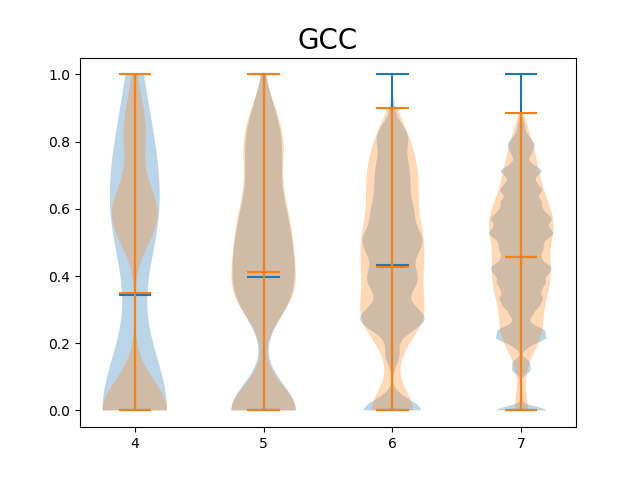}
	\includegraphics[width=.24\textwidth]{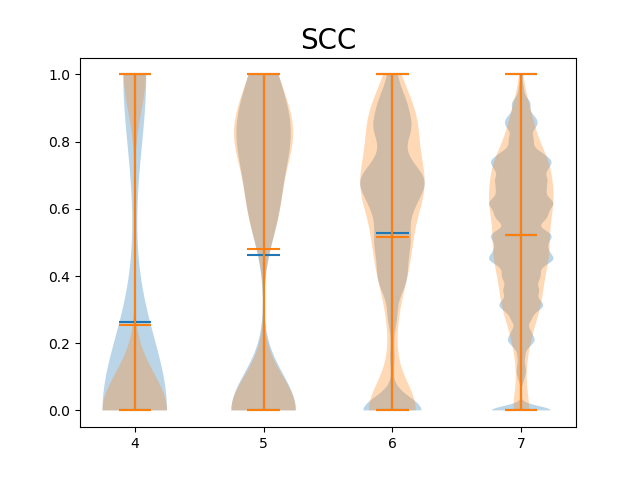}
    \includegraphics[width=.24\textwidth]{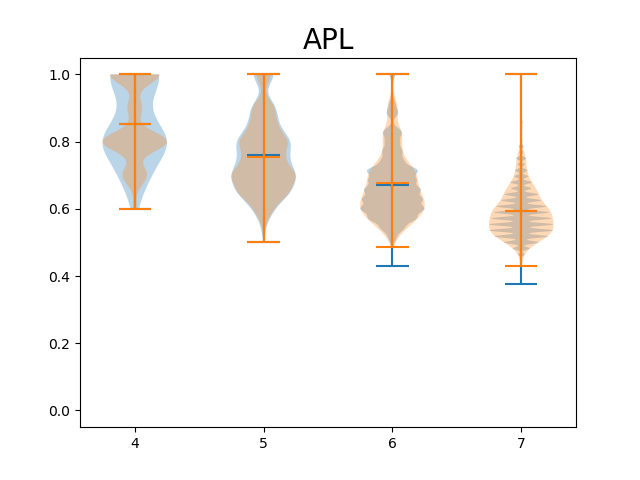}
	\includegraphics[width=.24\textwidth]{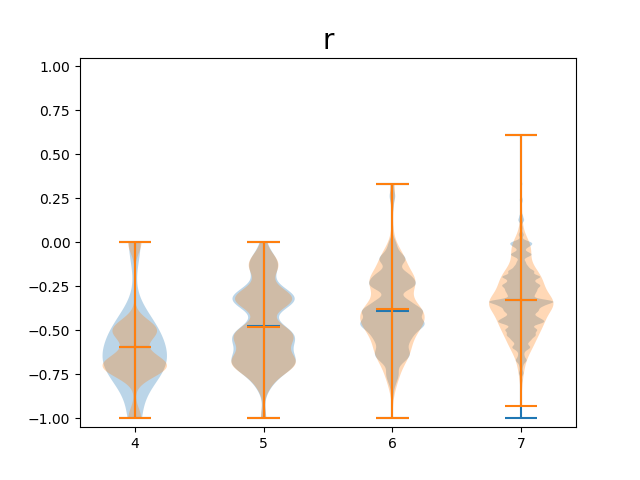}
	
	\includegraphics[width=.24\textwidth]{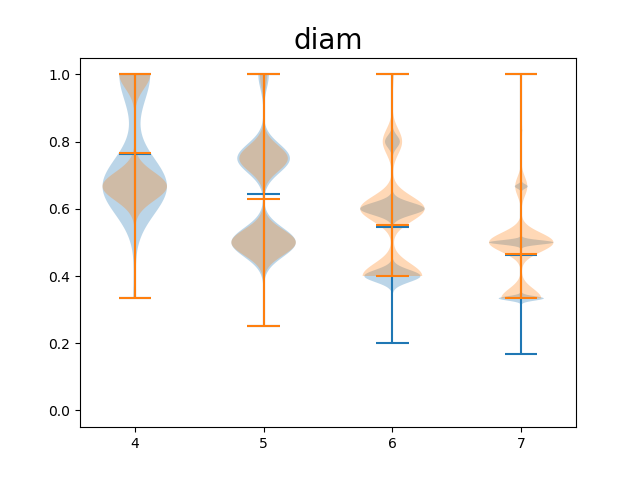}
	\includegraphics[width=.24\textwidth]{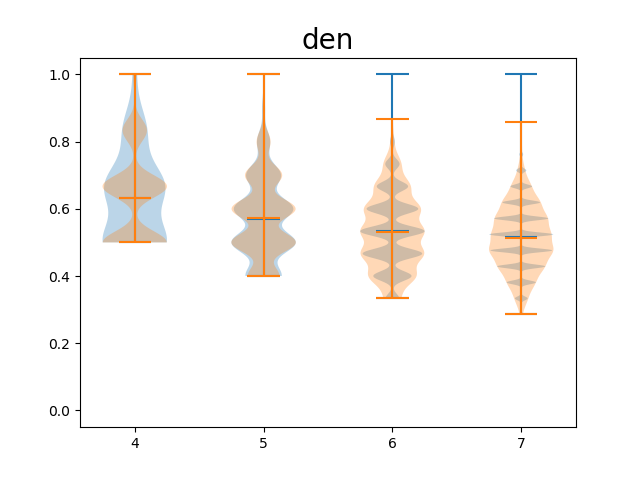}
	\includegraphics[width=.24\textwidth]{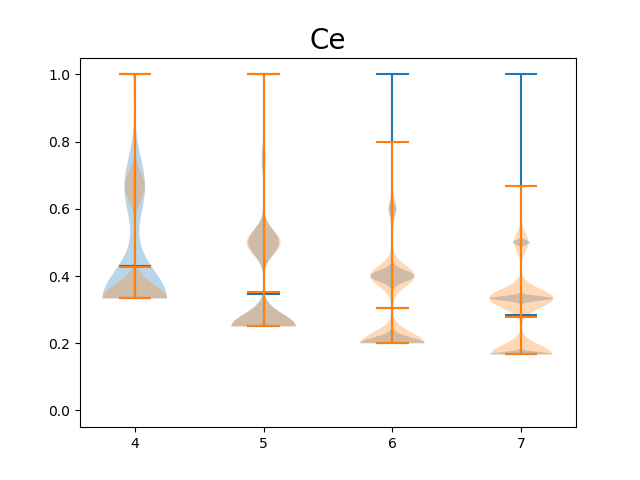}
	\includegraphics[width=.24\textwidth]{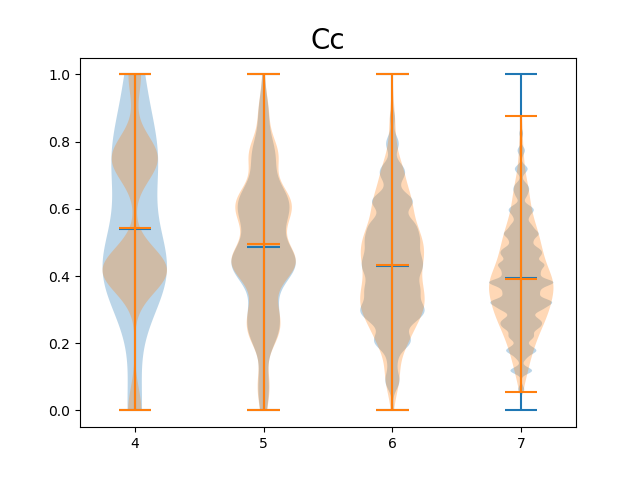}
	
	\includegraphics[width=.24\textwidth]{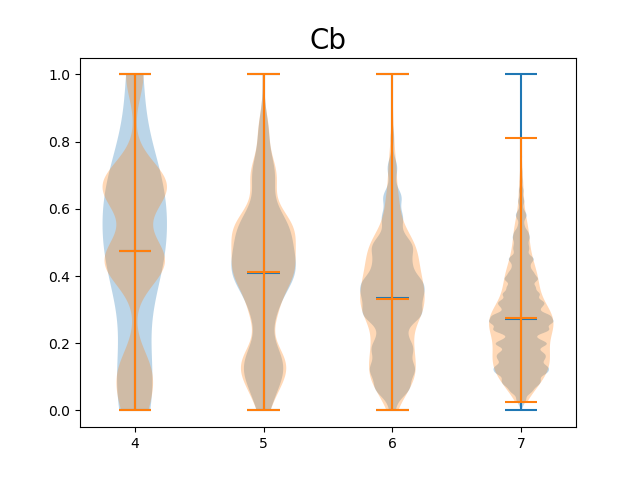}
    \includegraphics[width=.24\textwidth]{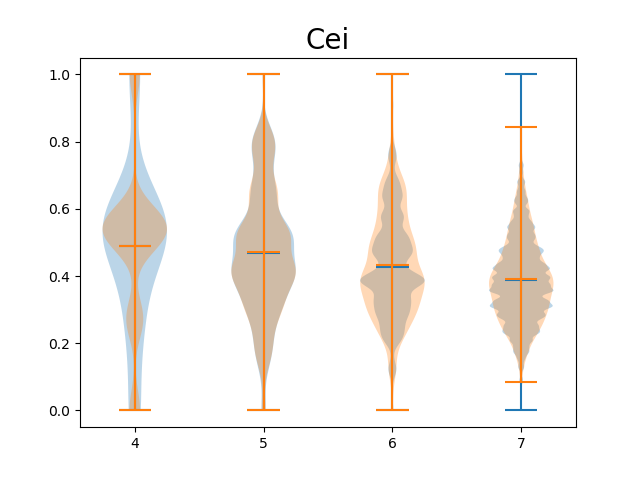}
    \includegraphics[width=.24\textwidth]{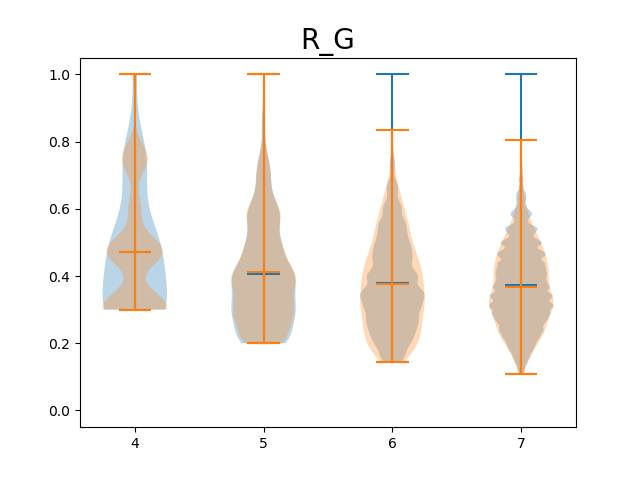}
    \includegraphics[width=.24\textwidth]{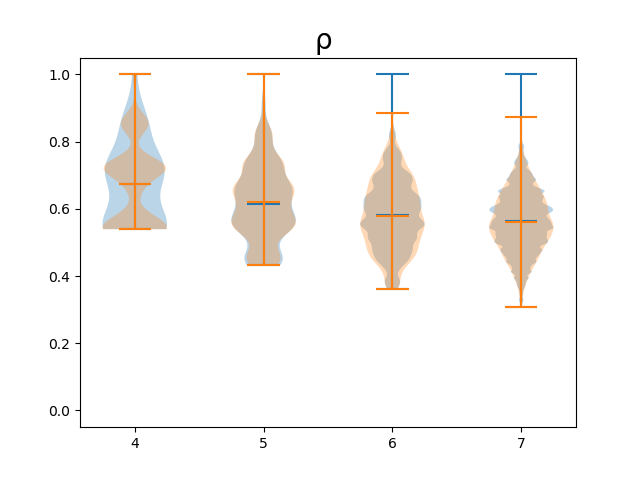}
	
	\caption{Distribution of the 12 graph properties for the set of all labeled graphs (see the blue violin plots of each subfigure) and $1000$ graphs generated by \ErdosRenyi model with $p = 1/2$ with values of $|V|$ in the range $[4,5,6,7]$ (see the orange violin plots of each subfigure). The first row demonstrates the results for Global Clustering Coefficient, Average Square Clustering, Average Path Length and Degree Assortativity.
	The second row demonstrates the results for Diameter, Density, Edge Connectivity and Closeness Centrality. The third row demonstrates the results for Betweenness Centrality, Eigenvector Centrality, Effective Graph Resistance and Spectral Radius.
	}
	\label{fig:violin_conv_3}
	\centering
\end{figure}

In Fig.~\ref{fig:violin_conv_3} we observe the distribution of graph properties for all labeled graphs with $|V|$ in the range $[4,5,6,7]$ and plot the distribution of each property in violin plots (see the blue violin plots of Fig.~\ref{fig:violin_conv_3}). Ideally, \ErdosRenyi with $p=1/2$ models well the space of all labeled graphs. Fig.~\ref{fig:violin_conv_3} demonstrates the correctness of this observation as the statistics for graphs generated by the \ErdosRenyi model with $p = 1/2$ matches well with the statistics of the set of all graphs (see the similarities of the blue and orange violin plots of Fig.~\ref{fig:violin_conv_3}).

\begin{figure}[hb!]
	\includegraphics[width=.24\textwidth]{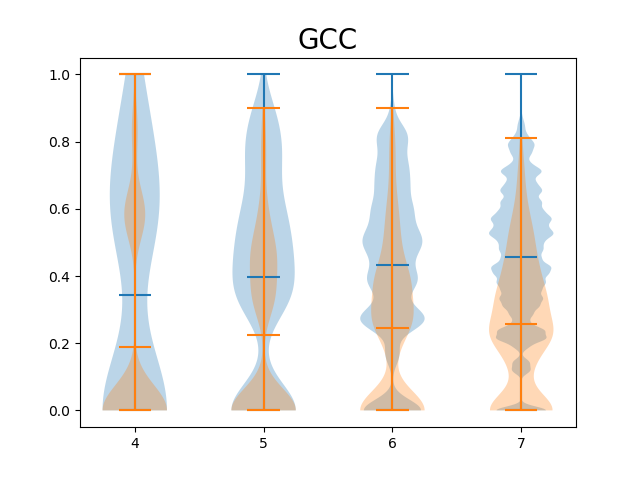}
	\includegraphics[width=.24\textwidth]{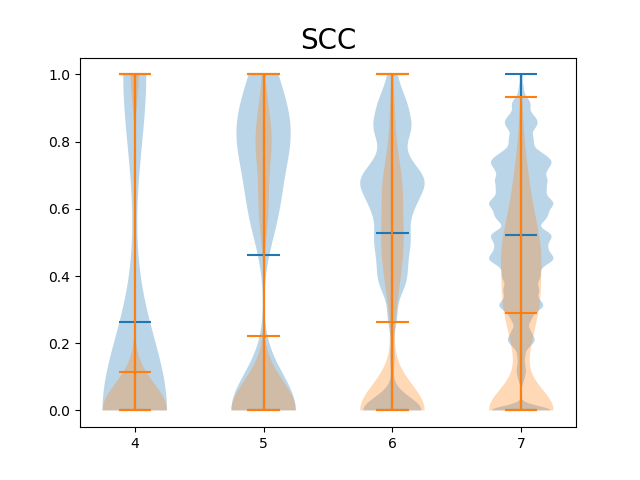}
    \includegraphics[width=.24\textwidth]{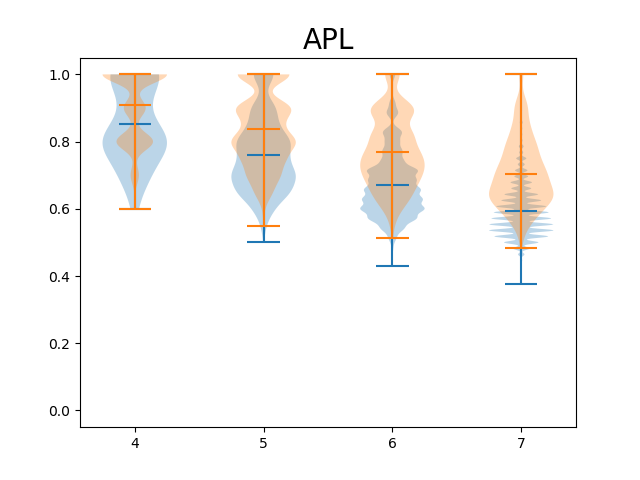}
	\includegraphics[width=.24\textwidth]{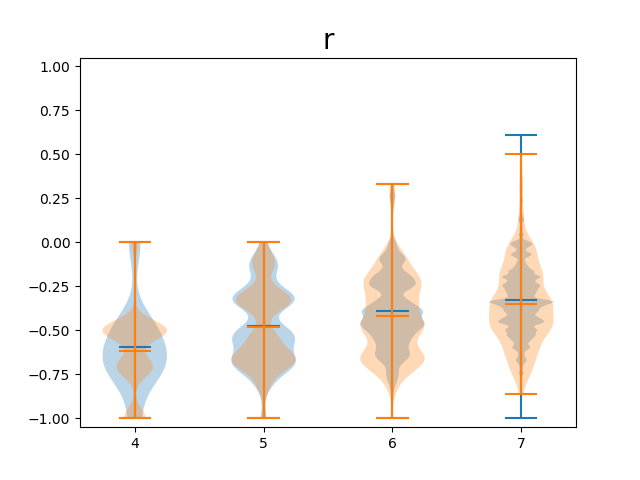}
	
	\includegraphics[width=.24\textwidth]{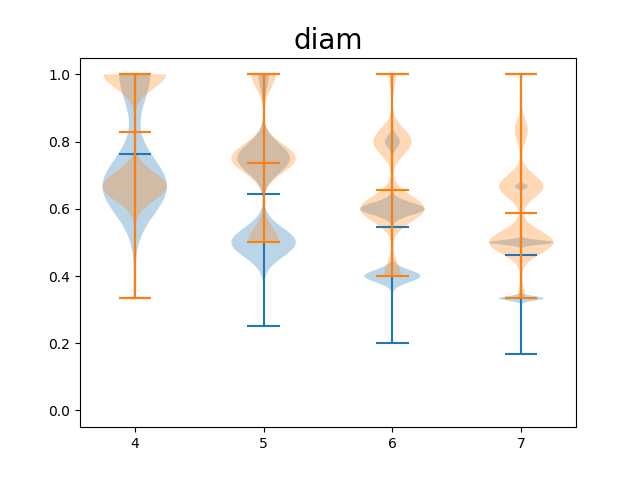}
	\includegraphics[width=.24\textwidth]{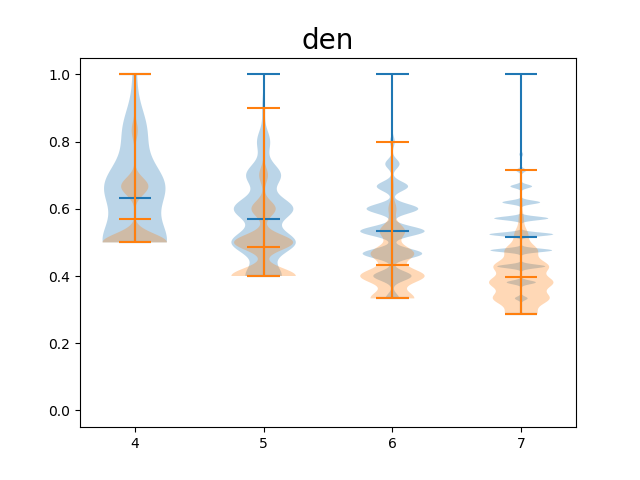}
	\includegraphics[width=.24\textwidth]{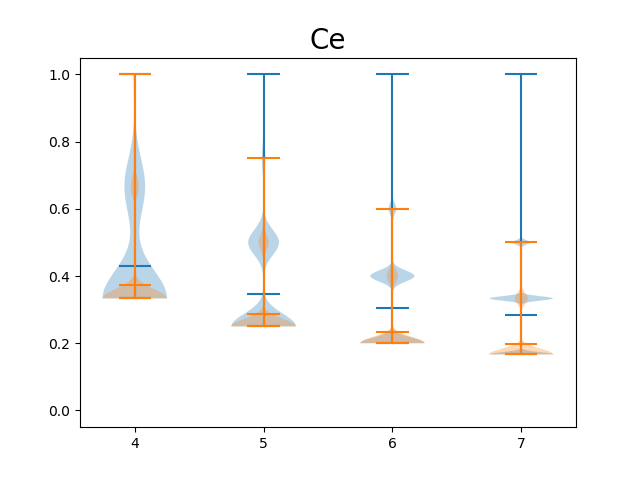}
	\includegraphics[width=.24\textwidth]{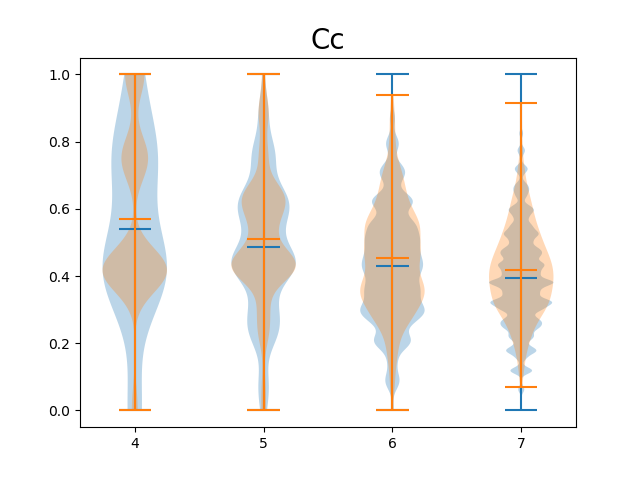}
	
	\includegraphics[width=.24\textwidth]{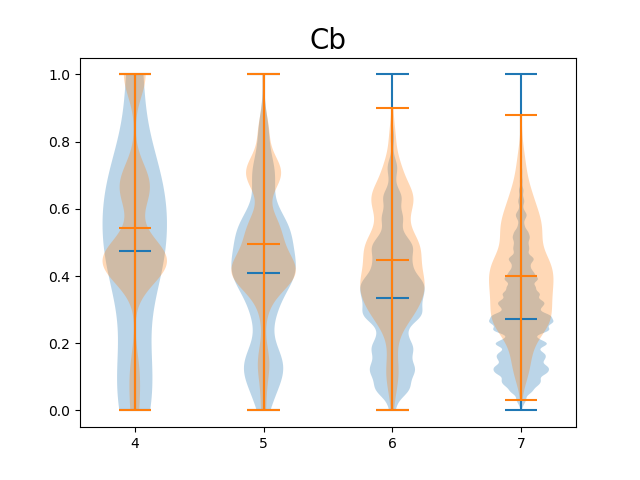}
    \includegraphics[width=.24\textwidth]{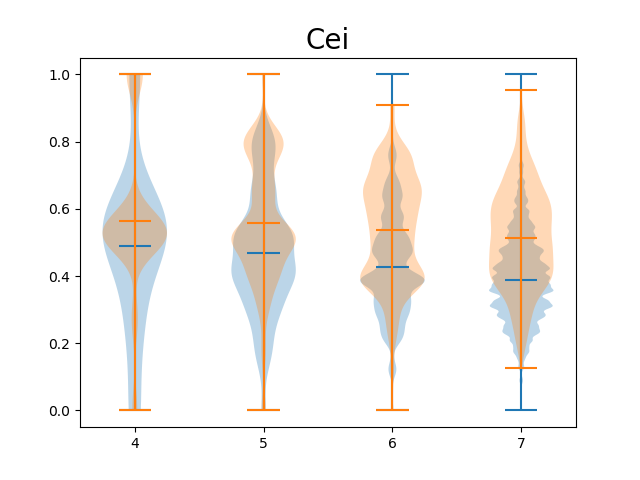}
    \includegraphics[width=.24\textwidth]{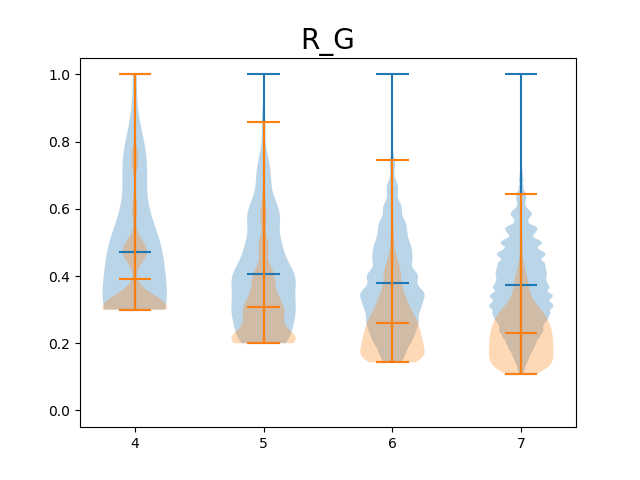}
    \includegraphics[width=.24\textwidth]{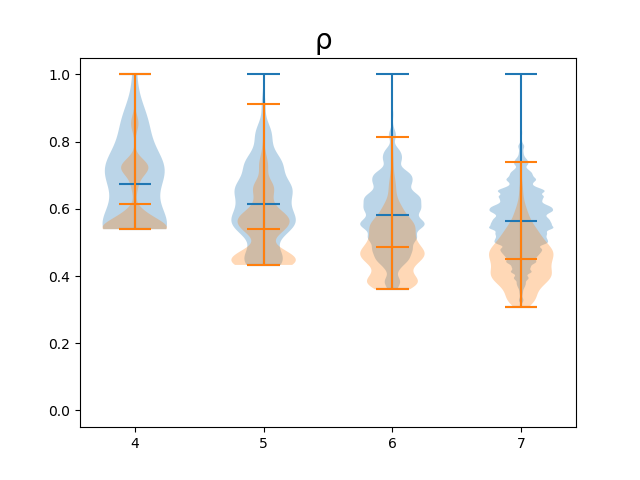}
	
    \caption{Distribution of the 12 graph properties for the set of all labeled graphs (see the blue violin plots of each subfigure) and $1000$ graphs generated by \ErdosRenyi model with $p = 1/3$ with values of $|V|$ in the range $[4,5,6,7]$ (see the orange violin plots of each subfigure). The first row demonstrates the results for Global Clustering Coefficient, Average Square Clustering, Average Path Length and Degree Assortativity.
	The second row demonstrates the results for Diameter, Density, Edge Connectivity and Closeness Centrality. The third row demonstrates the results for Betweenness Centrality, Eigenvector Centrality, Effective Graph Resistance and Spectral Radius.
	}
	\label{fig:violin_conv_4}
	\centering
\end{figure}

In Fig.~\ref{fig:violin_conv_4} we again observe the distribution of graph properties for all labeled graphs with $|V|$ in the range $[4,5,6,7]$ and plot the distribution of each property in violin plots (see the blue violin plots of Fig.~\ref{fig:violin_conv_4}. However, this time we compare it with 1000 graphs generated by the \ErdosRenyi model with $p = 1/3$. We notice that the mean values of the distributions are similar. However, the distributions of the properties are significantly different and does not match with the one from all labeled graphs.

\section{Classifying Between Graph Generators.}
\label{sec:graph_classications}

In this section we consider an application where the studied graph properties are used. We propose the following problem: Given graph generated by various random graph generators can we predict which generator actually generated the graph just by looking at the $15$ properties discussed in Sec.~\ref{sec:def_properties}. We believe that this application can be used for several purposes such as coming up with more complex random graph generators and graph anonymization.
We propose a model to classify between 8 different graph generators, using the $15$ graph properties. In addition to the \ErdosRenyi graph generator with $p = 1/2$ we also generate graphs based on the stochastic block model with $2, 3, 4$ and $5$ blocks, geometric (GE), Watts-Strogatz (WS) and  Barabasi-Albert (BA) graph generators.

The stochastic block model~\cite{holland1983stochastic}
generalizes the \ErdosRenyi model,  producing graphs that contain communities (clusters).
The parameters include the number of vertices $|V| = n$, the number of communities $l$ with a partition $C_1, C_2, \dots, C_l$ of $\{ 1, 2, \dots, n \}$, and a symmetric matrix $P \in \mathbb{R}^{l \times l}$ that determines the probabilities for adding edges within and between communities. For a pair of vertices $u$ and $v$ an edge is added with probability $P_{ij}$, where $u \in C_i$ and $v \in C_j$. 

The Watts-Strogatz~\cite{watts1998collective} (WS) model can be used to generate graphs that exhibit small-world properties (higher clustering coefficient and shorter average path lengths). We utilize the variation suggested by Newman and Watts~\cite{newman1999scaling} to ensure that the generated graphs are connected. 
The geometric model (GE)~\cite{gilbert1961random}, places nodes according to a Poisson point process in some metric space (e.g., the unit square in 2D), and adds edges between pairs of nodes that are within a pre-specified distance threshold. 
The Barabasi-Albert model (BA)~\cite{barabasi1999emergence} is a graph growth model where each added vertex has a fixed number of edges $|E|$, and the probability of each edge connecting to an existing vertex $v$ is proportional to the degree of $v$. 
%
\begin{table}[b!]
\caption{Demonstration of the top 10 pairs (left table) and triples (right table) of graph properties that achieve the best accuracy scores according to Random Forest.}
\vspace{0.3cm}
\begin{tabular}{|l|l|l|l|l|}
\hline
Properties                  & RF    & LR & SVM     & NN               \\ \hline
GCC,  $\rho$              & \textbf{89.2\%} & 55.5\%             & 89.2\% & 77.3\%          \\ \hline
GCC, ASCC                          & \textbf{85.4\%} & 55.4\%             & 65.8\% & 67.4\%          \\ \hline
GCC, den                          & \textbf{82.1\%} & 53.7\%             & 81.4\% & 72.3\%          \\ \hline
GCC, APL                          & \textbf{81.8\%} & 50.8\%             & 61.6\% & 64.5\%          \\ \hline
GCC, $R_G$   & \textbf{81.8\%} & 55.0\%             & 81.5\% & 79.7\%          \\ \hline
GCC, r                            & \textbf{78.2\%} & 72.7\%             & 79.6\% & 81.3\%          \\ \hline
GCC, $C_C$     & \textbf{77.0\%} & 62.0\%             & 74.0\% & 74.6\%          \\ \hline
GCC,  $C_{ei}$ & \textbf{75.5\%} & 56.9\%             & 76.0\% & 71.3\%          \\ \hline
GCC, $C_e$          & 73.5\%          & 63.5\%             & \textbf{76.8}\% & 75.8\% \\ \hline
GCC, $C_B$        & \textbf{71.3\%
} & 51.0\%             & 62.4\% & 69.6\%          \\ \hline
\end{tabular}
\quad
\begin{tabular}{|l|l|l|l|l|}
\hline
Properties                                   & RF    & LR & SVM     & NN      \\ \hline
\scalebox{0.75}[1]{GCC, ASCC, $\rho$   }                      & \textbf{92.8\%} & 56.6\%             & 72.4\% & 88.1\% \\ \hline
\scalebox{0.75}[1]{GCC, ASCC, APL}                                     & \textbf{92.5\%} & 54.3\%             & 66.1\% & 62.4\% \\ \hline
\scalebox{0.75}[1]{GCC, ASCC, den   }                                   & \textbf{92.5\%} & 56.7\%             & 72.4\% & 76.3\% \\ \hline
\scalebox{0.75}[1]{GCC, APL, $\rho$}                        & \textbf{92.0\%} & 55.2\%             & 70.6\% & 80.6\% \\ \hline
\scalebox{0.75}[1]{GCC, ASCC, $R_G$}               & \textbf{91.9\%} & 57.4\%             & 75.7\% & 73.5\% \\ \hline
\scalebox{0.75}[1]{GCC, den, $\rho$}                       & \textbf{91.7\%} & 54.1\%             & 91.2\% & 82.2\% \\ \hline
\scalebox{0.75}[1]{GCC, $C_e$, $\rho$}            & \textbf{90.9\%} & 63.9\%             & 82.3\% & 82.0\% \\ \hline
\scalebox{0.75}[1]{GCC, den, $R_G$  }           & \textbf{90.5\%} & 56.4\%             & 85.6\% & 75.5\% \\ \hline
\scalebox{0.75}[1]{GCC, r, $\rho$}                            & \textbf{90.4\%} & 72.6\%             & 79.9\% & 81.9\% \\ \hline
\scalebox{0.75}[1]{GCC, $C_{ei}$, $\rho$ }& \textbf{90.3\%} & 59.2\%             & 76.3\% & 81.3\% \\ \hline
\end{tabular}

\label{tbl:ml_accuracy}
\end{table}

We classify the graphs generated by the 8 models (\ErdosRenyi with $p = 1/2$, SBM with 2 blocks, SBM with 3 blocks, SBM with 4 blocks, SBM with 5 blocks, GE, WS and BA), using the $15$ graph properties. First, we generate 1000 graphs with each of the generators and calculate the $15$ properties for each of them. For the classification, we consider the graph properties as features and the generators as the labels. 
For the SBM we make sure that the blocks are of the same size and the symmetric matrix $P$ is randomly generated, such that, inside cluster probabilities (diagonal entries in $P$) have expected values equal to $0.75$. This condition ensures that the graphs are denser within the clusters than between the clusters. We randomly generate $20$ $P$ matrices for each of the SBM and use each of these $P$ matrices to generate $50$ graphs. Thus, for each SBM we generate 1000 graphs. We generate 1000 graphs with each of the other graph generators as well: ER with $p = 1/2$, GE, WS, BA. Thus, we have a total of 8000 graphs.

Since the generators have different density distributions and the range of many graph properties   depends on the density, we generate graphs that have densities around 0.5 (to match with those generated by \ErdosRenyi with $p = 1/2$).  Thus we fix the density for all generators in the range $den \in (0.47, 0.52)$. 

We use the following four standard classification algorithms with their default settings: random forest(RF), logistic regression(LR)~\cite{kleinbaum2002logistic}, kernel support vector machine (SVM)~\cite{keerthi2003asymptotic} and feed-forward neural network (NN)~\cite{svozil1997introduction} with $2$ hidden layers trained on the features. 
The choice of these classification algorithms is based on our experiments. From the list of many algorithms these showed the best results.
We test the four classification algorithms, on the generated dataset, which contains 8000 datapoints (graphs) with $15$ features (properties). 
To evaluate the effectiveness of these models and avoid overfitting we use 10-fold cross validation by splitting the data between training set (80\% of the total data) and testing set (20\% of the total data). 
The Random Forest (RF) achieves $91.8\%$ test accuracy, while the other three models have accuracy around $78\%$.
RF is also the most stable one over the 10 runs.

In Section~\ref{sec:property_relationships} we discussed the correlations between the $15$ graph properties. What we would like to test now is the following: In the classification task described above, can we achieve similar accuracy if instead of using all of the $15$ graph properties we use only a few of them?
As we have observed, graphs generated by the \ErdosRenyi model with $p = 1/2$ have four sets of correlated properties. The first set is: CC, APL, den, $R_G$ and $\rho$, the second one is: $C_C, C_B$ and $C_{ei}$, the third and fourth include only $r$ and $C_E$. 

Our next experiment is designed to see which pairs and triples of the $15$ graph properties can be most useful for the classification task. For this purpose we form all possible pairs and all possible triples from the $15$ properties and for each of them run the four classification algorithms (RF, LR, SVM and NN). To compute the accuracies, we again split the data of 8000 graphs between training set (80\% of the total dataset) and test set (20\% of the total dataset), compute the parameters of each model on training set and compute the achieved accuracies on test set. We repeat this procedure ten times and average the achieved accuracies. The results are reported in Table~\ref{tbl:ml_accuracy}. 
Note that some triples achieve a slightly higher accuracy than all $15$ properties, confirming 
the correlated property groups. 
Also notable is that GCC appears in all pairs and triples.
We believe that this is due to the clusters present in 4 of the 8 graph classes.

\begin{figure}[t]
\centering
    \includegraphics[width=0.42\textwidth]{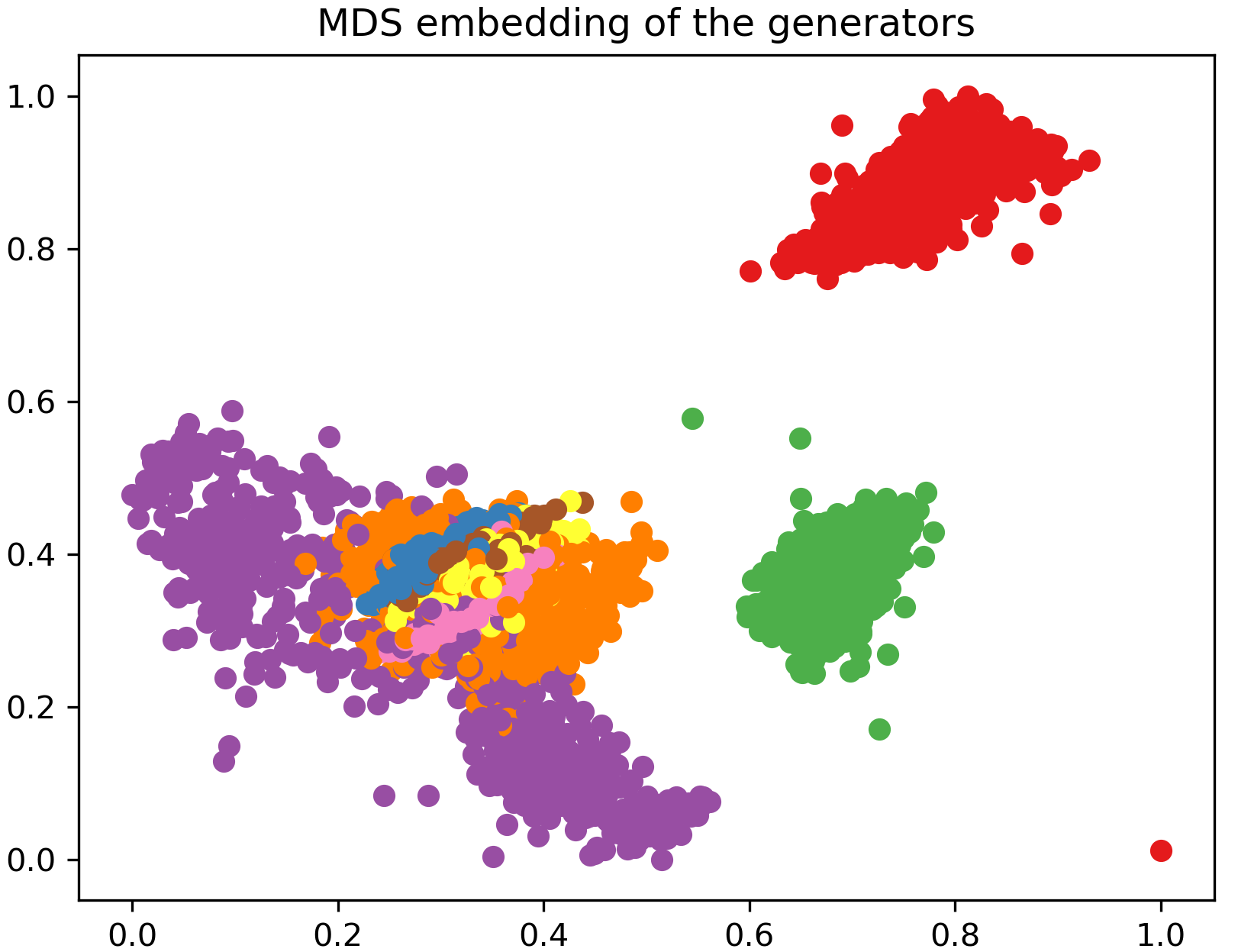}
    \includegraphics[width=0.14\textwidth]{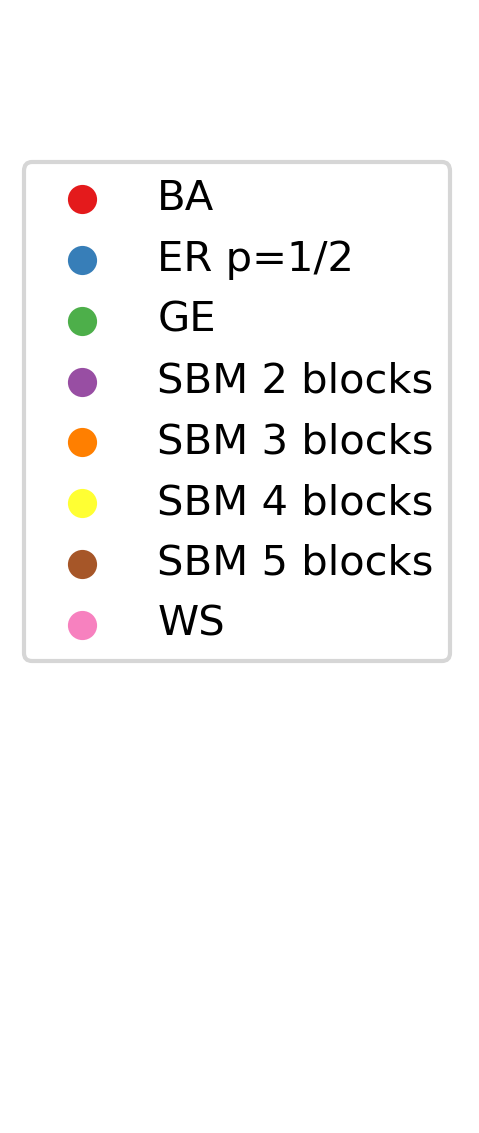}
    \includegraphics[width=0.42\textwidth]{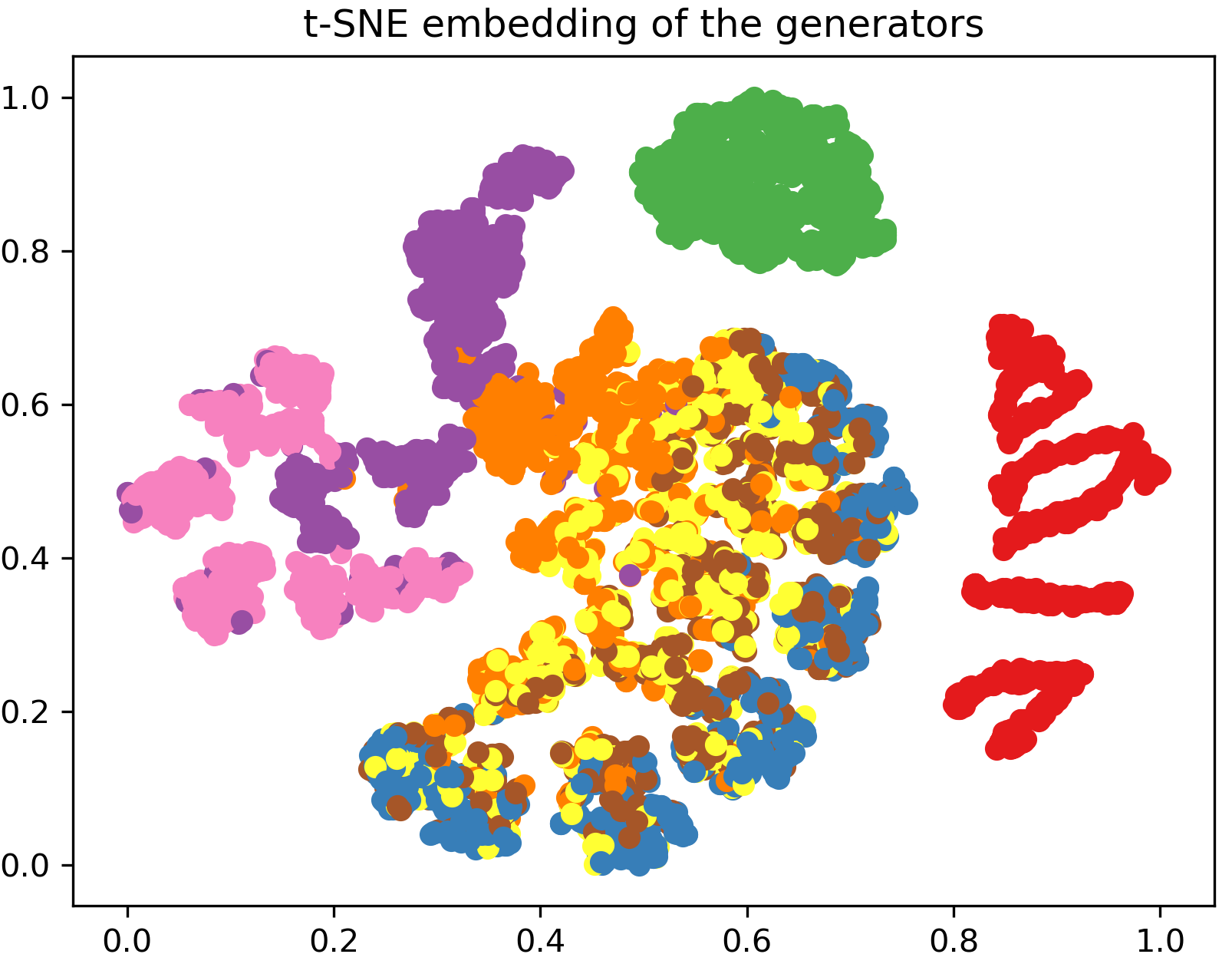}
\caption{Demonstration of the embedding of 8000 graphs generated by the 8 generators discussed in Section~\ref{sec:graph_classications} with $15$ graph properties in 2D.} 
  \label{fig:classification_embedding}
\end{figure}

Finally, we use MDS~\cite{mds} and t-SNE~\cite{t-sne} to visualize the 8000 graphs with computed $15$ properties in 2D; see Fig.~\ref{fig:classification_embedding}.  
We observe that both embeddings distinguish well between GE, WS and the rest in 2D, while the embeddings for \ErdosRenyi with 1/2 are mixed with the SBM with 2, 3, 4 and 5 blocks.

\begin{figure}[htp!]

	\includegraphics[width=.24\textwidth]{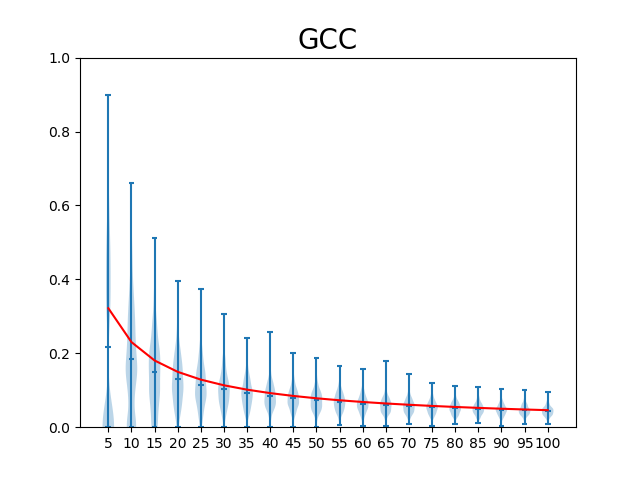}
	\includegraphics[width=.24\textwidth]{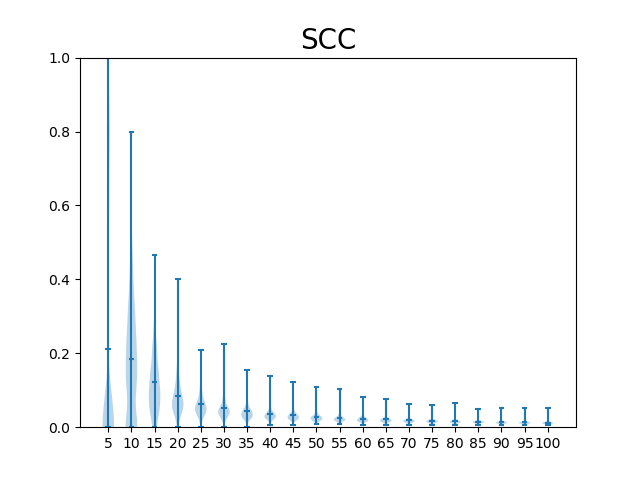}
	\includegraphics[width=.24\textwidth]{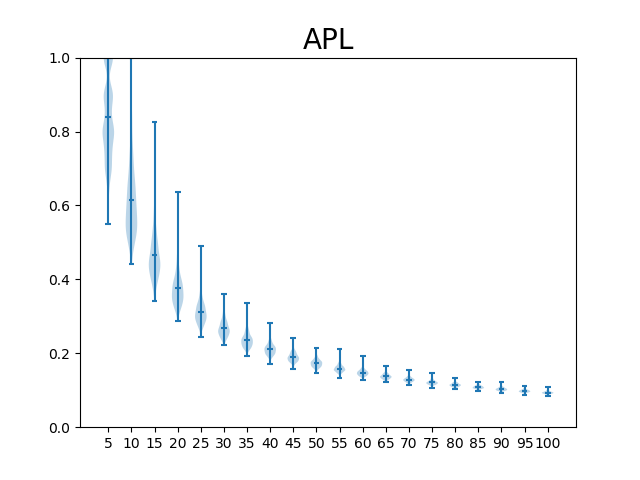}
    \includegraphics[width=.24\textwidth]{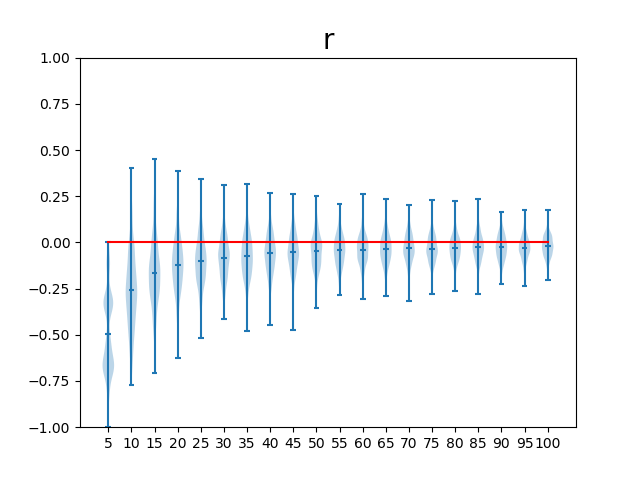}
    
    \includegraphics[width=.24\textwidth]{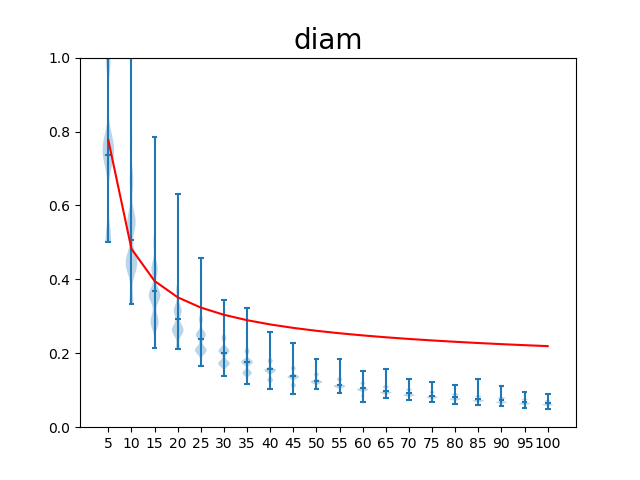}
	\includegraphics[width=.24\textwidth]{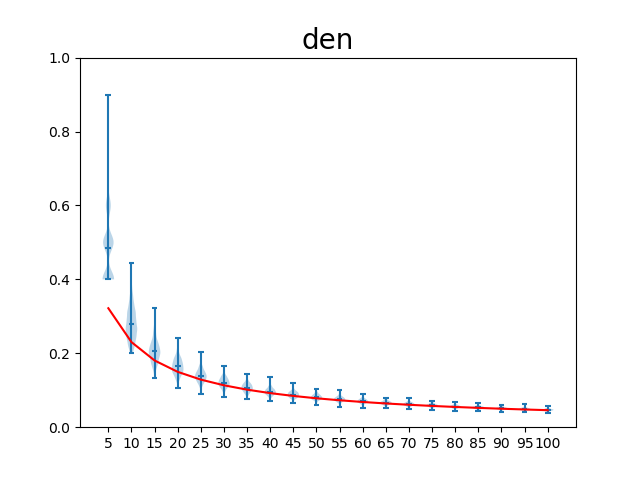}
	\includegraphics[width=.24\textwidth]{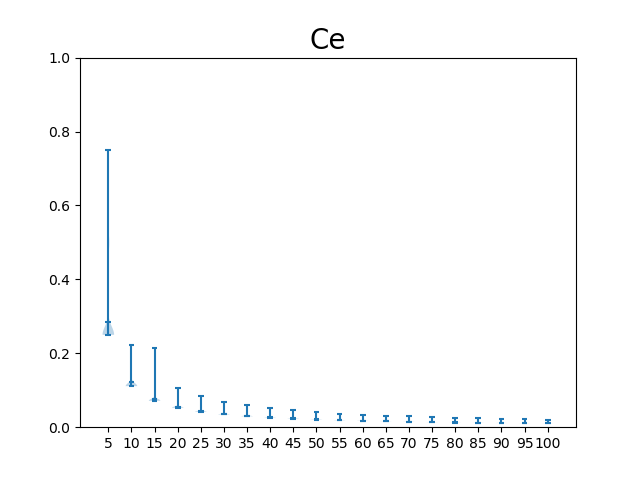}
	\includegraphics[width=.24\textwidth]{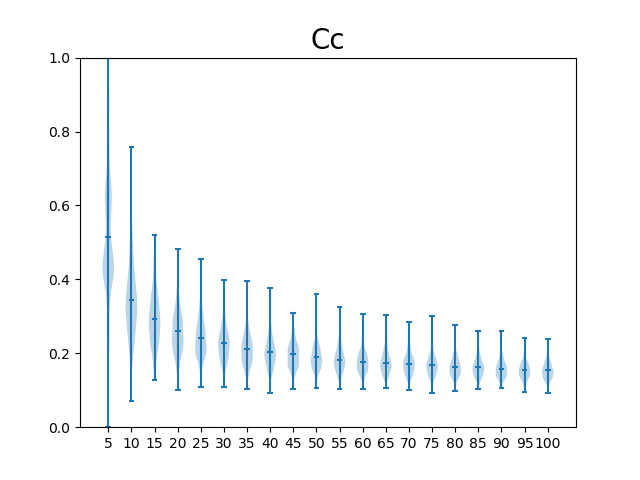}
	
	\includegraphics[width=.24\textwidth]{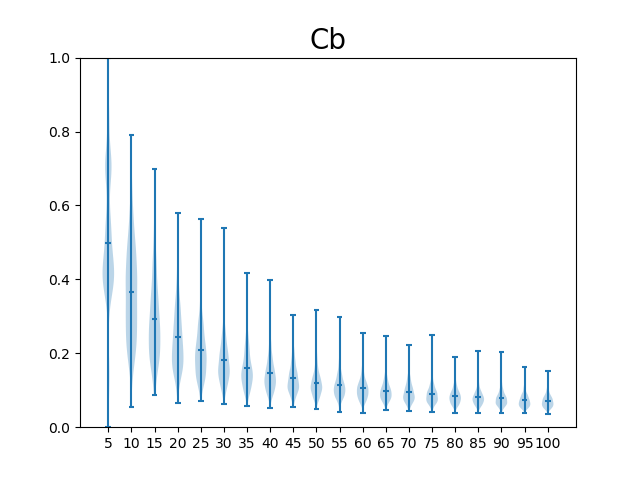}
	\includegraphics[width=.24\textwidth]{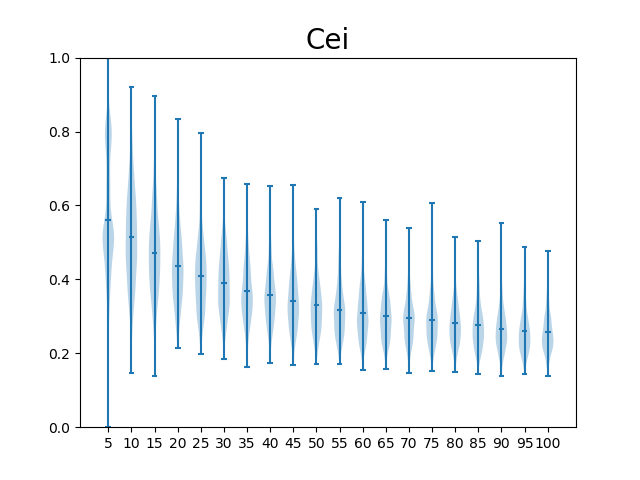}
	\includegraphics[width=.24\textwidth]{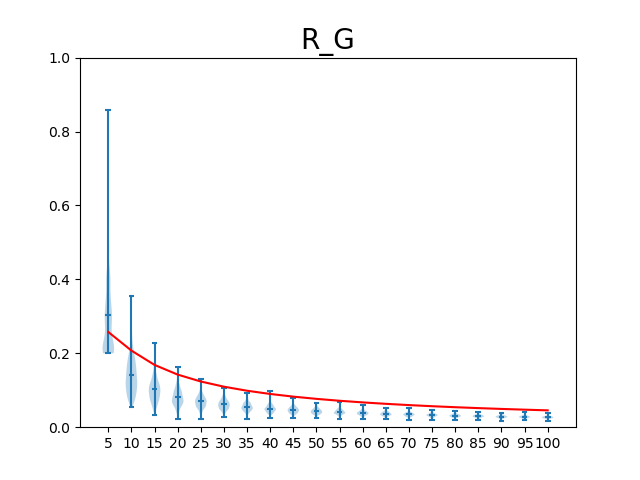}
	\includegraphics[width=.24\textwidth]{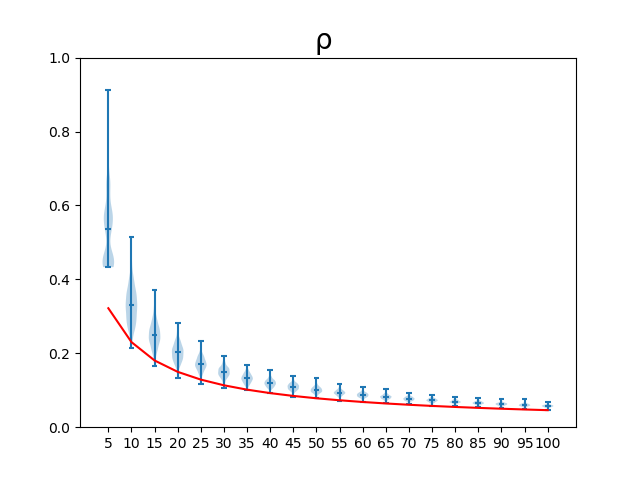}
	
	\includegraphics[width=.24\textwidth]{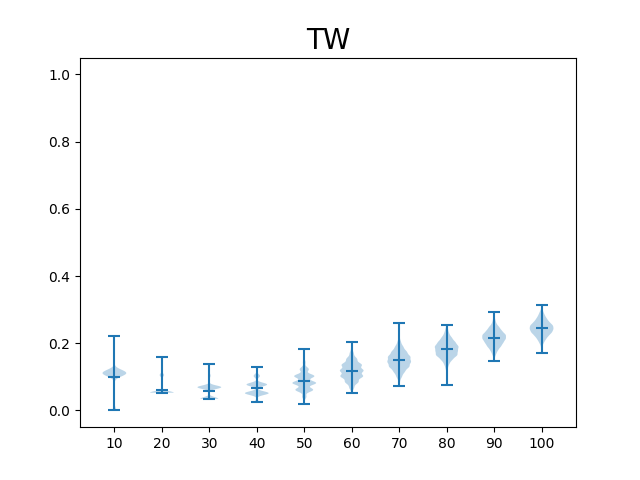}

\caption{Behavior of graph properties for $1000$ graphs generated according to \ErdosRenyi model with $p = \log(|V|) / |V|$. The number of vertices is in the range $[5,10,..,100]$ and the results are illustrated with violin plots. The first row shows the results for Global Clustering Coefficient, Average Square Clustering, Average Path Length and Degree Assortativity.
The second row includes Diameter, Density, Edge Connectivity and Closeness Centrality. The third row includes Betweenness Centrality, Eigenvector Centrality, Effective Graph Resistance and Spectral Radius. The fourth row includes the results for Treewidth, TreeDepth and Vertex Covering.
}
	\label{fig:violin_conv_log}
	\centering
\end{figure}

\end{document}